\documentclass[aps,prc,reprint,superscriptaddress,nofootinbib,floatfix,showpacs]{revtex4-1}
\pdfoutput=1

\usepackage{epsfig}
\usepackage{dcolumn}
\usepackage{bm}
\usepackage{amssymb}
\usepackage{amsmath}
\usepackage{hyperref}

\begin{document}
\title{Effect of the QCD equation of state and strange hadronic resonances on multiparticle correlations in heavy ion collisions}

\author{P. Alba}
\affiliation{ Frankfurt Institute for Advanced Studies,
Goethe Universit\"at Frankfurt, D-60438 Frankfurt am Main, Germany}
\author{V. Mantovani  Sarti}
\affiliation{Physik Department T70, E62, Technische Universit\"at M\"unchen,
James Franck Strasse 1, 85748 Garching, Germany}
\author{J. Noronha}
\affiliation{Instituto de F\'isica, Universidade de S\~ao Paulo, Rua do Mat\~ao
1371, Butant\~a, 05508-090, S\~ao Paulo, SP, Brazil}
\author{J. Noronha-Hostler}
\affiliation{Department of Physics and Astronomy, Rutgers University, Piscataway, NJ USA 08854}
\author{P. Parotto}
\affiliation{Department of Physics, University of Houston, Houston, TX, USA  77204}
\author{I. Portillo  Vazquez}
\affiliation{Department of Physics, University of Houston, Houston, TX, USA  77204}
\author{C. Ratti}
\affiliation{Department of Physics, University of Houston, Houston, TX, USA  77204}

\date{\today}
\begin{abstract}
The QCD equation of state at zero baryon chemical potential is the only element of the standard dynamical framework to describe heavy ion collisions that can be directly determined from first principles. Continuum extrapolated lattice QCD equations of state have been computed using 2+1 quark flavors (up/down and strange) as well as 2+1+1 flavors to investigate the effect of thermalized charm quarks on QCD thermodynamics. Lattice results have also indicated the presence of new strange resonances that not only contribute to the equation of state of QCD matter but also affect hadronic afterburners used to model the later stages of heavy ion collisions. We investigate how these new developments obtained from first principles calculations affect multiparticle correlations in heavy ion collisions. We compare the commonly used equation of state S95n-v1, which was constructed using what are now considered outdated lattice results and hadron states, to the current state-of-the-art lattice QCD equations of state with 2+1 and 2+1+1 flavors coupled to the most up-to-date hadronic resonances and their decays. New hadronic resonances lead to an enhancement in the hadronic spectra at intermediate $p_T$. Using an outdated equation of state can directly affect the extraction of the shear viscosity to entropy density ratio, $\eta/s$, of the quark-gluon plasma and results for different flow observables. The effects of the QCD equation of state on multiparticle correlations of identified particles are determined for both AuAu $\sqrt{s_{NN}}=200$ GeV and PbPb $\sqrt{s_{NN}}=5.02$ TeV collisions. New insights into the $v_2\{2\}$ to $v_3\{2\}$ puzzle in ultracentral collisions are found. Flow observables of heavier particles exhibit more non-linear behavior regardless of the assumptions about the equation of state, which may provide a new way to constrain the temperature dependence of $\eta/s$.

\end{abstract}

\maketitle

\section{Introduction}

Relativistic heavy ion collisions at the Relativistic Heavy Ion Collider (RHIC) and the Large Hadron Collider (LHC) have successfully recreated the Quark-Gluon Plasma (QGP), an exotic state of matter predicted by Quantum Chromodynamics (QCD) to have existed in the early Universe where quarks and gluons are not confined into hadrons. In thermodynamic equilibrium (and at zero baryon chemical potential), the interactions between quarks and gluons in the QGP can be computed from first principles using lattice field theory techniques \cite{Gattringer:2010zz}, i.e., Lattice QCD. This approach has shown that the QCD phase transition at zero baryon chemical potential is a smooth crossover \cite{Aoki:2006we} and the full result for the QCD Equation of State (EoS) with 2+1 quark flavors is now known \cite{Borsanyi:2013bia,Bazavov:2014pvz}. More recently, in \cite{Borsanyi:2016ksw} calculations for the QCD EoS have been extended to include also the effects from thermalized charm quarks, which provides an interesting opportunity to probe how the active flavor content of the QGP in equilibrium affects dynamical observables computed via hydrodynamic simulations.

Given that lattice results for the QCD EoS are generally not available at low temperatures, in practice to obtain an EoS that can be used in hydrodynamic simulations across a large range of temperatures the lattice part of the EoS is matched to a Hadron Resonance Gas (HRG) model at low temperatures $T < 130 $ MeV.  In its simplest form (i.e., a non-interacting gas of hadrons and their resonances) \cite{Karsch:2003zq}, the main variable in the HRG model is the spectrum of hadronic states obtained from the Particle Data Group (PDG).  In the past there was a mismatch at low temperatures between lattice thermodynamic results and calculations from HRG, which was understood as a consequence of the use of coarse lattices and the inclusion of too few particles in the hadronic spectrum  \cite{Eidelman:2004wy}. As lattice calculations became more refined, previous works overcame this issue by adjusting the masses of the hadron resonance gas \cite{Karsch:2003zq,Huovinen:2009yb} or by implementing volume corrections (see, for instance, \cite{NoronhaHostler:2009cf}). Another possibility was that there could be yet undiscovered resonances that could be added to the hadron spectrum \cite{NoronhaHostler:2009cf,Majumder:2010ik,NoronhaHostler:2012ug}\footnote{The addition of extra resonances would also affect other quantities relevant for heavy ion collision modeling such as the temperature dependence of $\eta/s$ and $\zeta/s$ (with $\zeta$ being the bulk viscosity) in the hadronic phase \cite{NoronhaHostler:2008ju,NoronhaHostler:2012ug} (leading to a dip in $\eta/s$ and a peak in $\zeta/s$ near the crossover). Extra resonances were also found to suppress elliptic flow at intermediate $p_T$ \cite{Noronha-Hostler:2013rcw}, mildly improve the $\chi^2$ of thermal fits \cite{NoronhaHostler:2009tz}, and allow for chemical equilibrium in the hadron gas phase to be reached dynamically on very short time scales \cite{NoronhaHostler:2007jf,NoronhaHostler:2009cf,Noronha-Hostler:2014usa,Noronha-Hostler:2014aia}.}. 
During this time period there were also large discrepancies between the results for the QCD EoS obtained by different lattice groups \cite{Aoki:2006br,Cheng:2006qk}, which have converged in the last three years to the final answer in the case of 2+1 flavors \cite{Borsanyi:2013bia,Bazavov:2014pvz}. Concurrently, PDG added new states \cite{Agashe:2014kda} (the ones that are most experimentally certain indicated by ***-****, totaling at about $\sim 300$ resonances), which improved the HRG result for the pressure making it possible to match it to lattice calculations at $T\sim 155$ MeV  \cite{Borsanyi:2013bia,Bazavov:2014pvz}. However, more sensitive lattice QCD observables such as the susceptibilities of conserved charges, the ratio between the baryon and strangeness chemical potentials $\mu_B/\mu_S$ \cite{Bazavov:2014xya}, and also the partial pressures \cite{Alba:2017mqu}, still indicate the need for additional strange states beyond those currently used in heavy ion collisions modeling.  In fact, using the most uncertain PDG resonances classified as *-**** \cite{Patrignani:2016xqp} the  HRG results can match the more differential lattice QCD calculations reasonably well, as shown in \cite{Alba:2017mqu}.  When using all *-**** states from the PDG we will refer to this cocktail of resonances (and their corresponding decays) as PDG16+ where the + is used to indicate the addition of the *-** states that have been traditionally excluded from HRG calculations.

The EoS is one of the inputs used in event-by-event viscous hydrodynamical simulations of the QGP. Following the hydrodynamic evolution, hadronic interactions are taken into account using hadronic afterburners (such as \cite{Bleicher:1999xi,Kolb:2002ve,Petersen:2008dd}). However, in the simple scenario pursued here no additional hadronic interactions are considered and, thus, after the hydrodynamic evolution and corresponding freeze-out process only the results from hadronic decays are implemented. Nevertheless, in this paper we include for the first time in hydrodynamic simulations the effects of all the listed strong decays of *-**** states with their corresponding non negligible branching ratios ($\simeq 1\%$ or higher) from \cite{Patrignani:2016xqp}, discarding weak decay channels.

Now that lattice calculations are able to determine the QCD equation of state in the continuum limit, new questions regarding the nature of the QCD crossover phase transition can be investigated.  For instance, in \cite{Borsanyi:2016ksw} it was shown that the inclusion of the charm quark in the QCD EoS could significantly alter the trace anomaly at temperatures above $T>300$ MeV, which are reachable by both RHIC and LHC. Though it is not clear if charm quarks are indeed thermalized within the dynamical QGP formed in heavy ion collisions, one can now for the first time check how the inclusion of an additional quark flavor in the EoS changes different sets of flow observables and get insight into the question concerning charm thermalization via direct comparisons of hydrodynamic calculations to experimental data.

The influence of the equation of state in heavy ion collisions has been recently studied in \cite{Dudek:2014qoa,Pratt:2015zsa,Moreland:2015dvc,Nara:2016hbg,Monnai:2017cbv}. The EoS used in most of the current hydrodynamic simulations was developed in Ref.\ \cite{Huovinen:2009yb}, which represented a significant advance in field at the time by combining the available lattice QCD results with detailed HRG calculations based on PDG 2005. However, this construction used lattice QCD results that were not extrapolated to the continuum limit and also an old version of the list of hadronic resonances that we now understand was missing a large number of states. In this paper we investigate how improvements on the QCD EoS coming from both the lattice perspective (i.e., results in the continuum limit) as well as from the most updated list of resonances in the HRG affect a large number of flow observables and the corresponding estimate for $\eta/s$. Furthermore, we seek to find observables that are sensitive to the inclusion of thermalized charm quarks within the equation of state.  In order to conduct this investigation, we use the event-by-event viscous hydrodynamic model, v-USPhydro \cite{Noronha-Hostler:2013gga,Noronha-Hostler:2014dqa,Noronha-Hostler:2015coa}, combined with resonance decays and make comparisons to experimental data at RHIC and LHC run 2. Our investigations lead to a number of relevant results:
\begin{itemize}
\item Updated resonance decays produce more particles at high $p_T$, which increases the $\langle p_T\rangle$ (especially for kaons and protons).
\item There is a direct connection between the equation of state and the extraction of the shear viscosity.  Changing from the previous EoS \cite{Huovinen:2009yb} to the new EoS presented here that includes state-of-the-art lattice QCD results in the continuum limit and an up-to-date list of hadronic resonances can alter the estimate of $\eta/s$ by nearly 50\%. 
\item The ratio $v_2\{2\}/v_3\{2\}$ in ultracentral collisions varies with the center of mass collision energy. Using the updated equation of state brings this ratio closer to experimental data but for $0-1\%$ centralities at LHC run 2 the model calculations are still $\sim 15-20\%$ above the data. 
\item We make predictions for multiparticle cumulants at LHC run 2.
\item We find that the linear mapping between eccentricities onto the final flow harmonics is strongly dependent on the mass of the observed particle.  Thus, heavier particles appear to have a larger contribution from non-linear effects in this context.  This effect is independent of the equation of state.
\item Symmetric cumulants and multiparticle cumulants of identified particles are studied for the first time.  We find that $v_2\{4\}/v_2\{2\}$ has no mass dependence whereas symmetric cumulants scale depending on the identified particle being used to calculate them. 
\item We study how $SC(4,2)$ scales from RHIC AuAu $\sqrt{s_{NN}}=200$ GeV collisions to LHC PbPb $\sqrt{s_{NN}}=5.02$ TeV.  We find that non-linear effects play a larger role at RHIC, which leads to a larger $SC(4,2)$.  
\end{itemize}

This paper is organized as follows.  In Section \ref{sec:fit} we explain how the new EoS is constructed using the most recent lattice QCD results from the Wuppertal-Budapest (WB) collaboration and HRG calculations performed with the most updated particle list PDG16+.  In Section \ref{sec:hydro} the hydrodynamic model is described and the extraction of $\eta/s$ using the most common flow observables is carried out. In Section \ref{sec:map} we show our results for flow harmonics, symmetric cumulants, and Pearson coefficients, respectively.  Finally, in Section \ref{sec:con} our conclusions and outlook are presented.

\section{Construction of new QCD EoS for a wide range of temperatures}\label{sec:fit}

The new equations of state presented here are composed of three separate pieces depending on the temperature: when $T>153$ MeV we use the WB Collaboration lattice QCD 2+1(+1) fitted EoS \cite{Borsanyi:2013bia,Borsanyi:2016ksw}, when $33.5 < T $[MeV]$ < 153 $ an HRG model with all known PDG resonances (including the contribution from *-**** states) is employed, while below $T< 33.5$ MeV a pion gas is used.  In order to connect them to produce the final trace anomaly a smoothing function ($\tanh$) is employed. The trace anomaly is given by
\begin{widetext}
\begin{equation}
\left(\frac{\varepsilon-3p}{T^4}\right)_{all}=\left(\frac{\varepsilon-3p}{T^4}\right)_{\pi}-\frac{1+\tanh\left[b(T-T_{HRG+Latt/\pi}) \right] }{2}\left[\left(\frac{\varepsilon-3p}{T^4}\right)_{HRG+Latt}-\left(\frac{\varepsilon-3p}{T^4}\right)_{\pi}\right]
\end{equation}
\end{widetext}
where $b=1$ MeV$^{-1}$ and $T_{HRG+Latt/\pi}=33.5$ MeV. For temperatures above $T>250$ MeV only the function fitted to the lattice QCD EoS from \cite{Borsanyi:2016ksw} is taken to avoid numerical issues with an HRG model at high temperatures. In the expression above, besides the pion gas contribution to the trace anomaly we have also defined
\begin{widetext}
\begin{equation}
\left(\frac{\varepsilon-3p}{T^4}\right)_{HRG+Latt}=\left(\frac{\varepsilon-3p}{T^4}\right)_{HRG}-\frac{1+\tanh\left[a(T-T_{HRG/Latt}) \right] }{2}\left[\left(\frac{\varepsilon-3p}{T^4}\right)_{Latt}-\left(\frac{\varepsilon-3p}{T^4}\right)_{HRG}\right]
\end{equation}
\end{widetext}
where $a=0.1$ MeV$^{-1}$ and $T_{HRG/Latt}=153$ MeV is the temperature chosen in such a way that the hadron resonance gas and the lattice results have good agreement. Above, the lattice input corresponds to the fitted function presented in \cite{Borsanyi:2016ksw}. 

\begin{figure}[ht]
\includegraphics[width=0.5\textwidth]{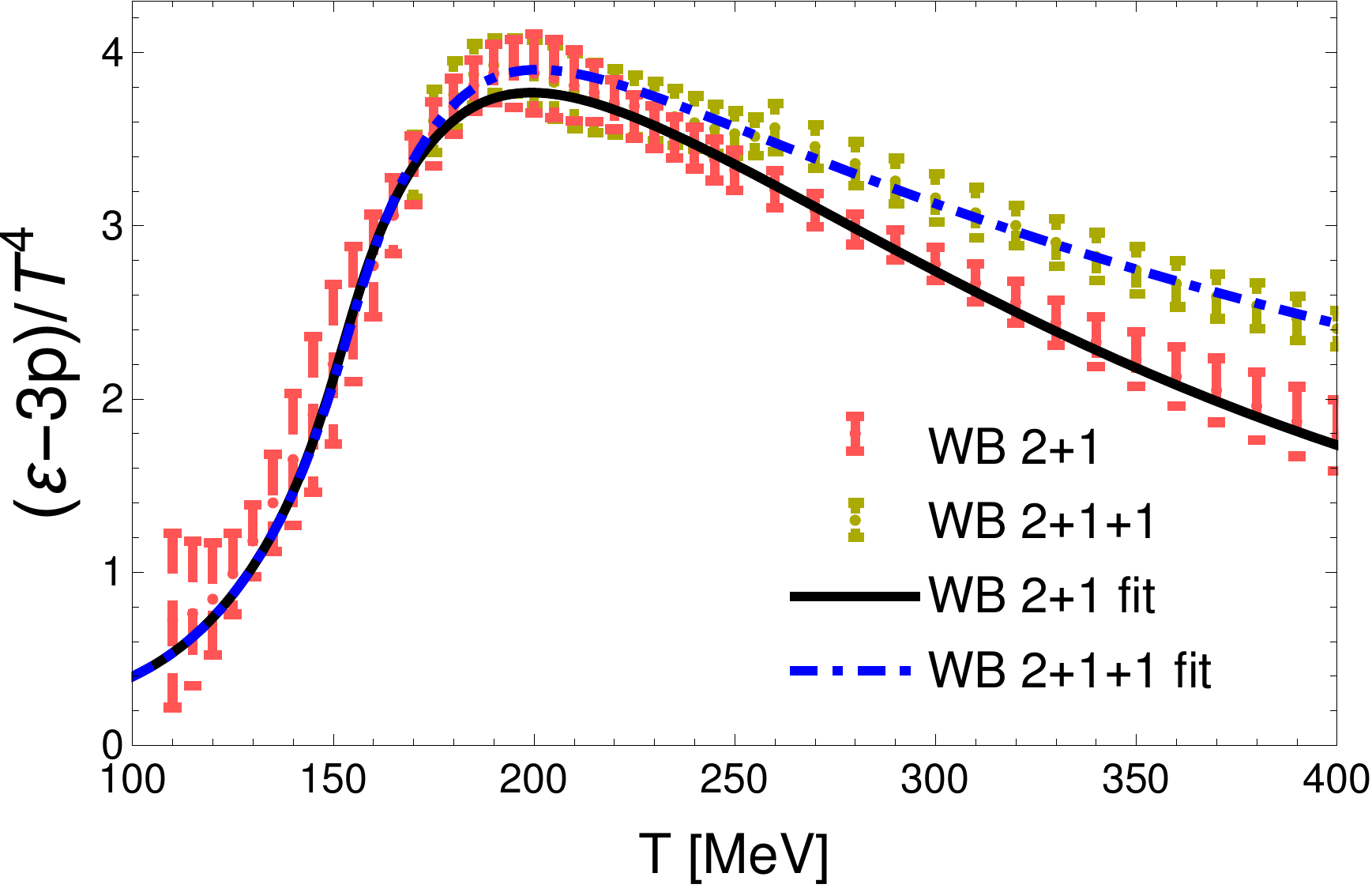} 
\caption{(Color online) Trace anomaly of QCD with 2+1 and 2+1+1 flavors computed by the  Wuppertal Budapest Collaboration \cite{Borsanyi:2013bia,Borsanyi:2016ksw} compared to the corresponding EoS constructed here.}
\label{fig:fitting}
\end{figure}

From the trace anomaly all other thermodynamic relations can be computed at $\mu_B=0$.  The pressure is obtained using:
\begin{equation*}
\frac{p}{T^4}=\int_0^{T}dT \;\frac{1}{T}\left(\frac{\varepsilon-3p}{T^4}\right)_{all}
\end{equation*}
and the rest follow such that:
\begin{eqnarray}
\frac{\varepsilon}{T^4}&=&\left(\frac{\varepsilon-3p}{T^4}\right)_{all}+\frac{3p}{T^4} \\
\frac{s}{T^3}&=&\left(\frac{\varepsilon-3p}{T^4}\right)_{all}+\frac{4p}{T^4} \\
c_s^2&=&\frac{s}{T}\frac{dT}{ds}=\frac{dp}{d\varepsilon} \\
\end{eqnarray}
In Fig.\ \ref{fig:fitting} we compare the lattice results for the QCD trace anomaly of 2+1 and 2+1+1 flavors with the reconstructed EoS discussed here. Using  PDG16+ with all *-**** resonances the best agreement between the two types of equations of state is found around $T\sim150-160$ MeV.  Note that for lower temperatures the lattice calculations in  \cite{Borsanyi:2016ksw} have large error bars and the HRG results computed using PDG16+ are well within those error bars until the lowest available temperatures $T\sim 100$ MeV. 

When Ref.\ \cite{Huovinen:2009yb}  was written relativistic hydrodynamical models still solved ideal hydrodynamics equations of motion and partial chemical equilibrium was implemented. Since then, with the advent of viscous hydrodynamics and its subsequent coupling to hadronic cascade models (as in hybrid models \cite{Petersen:2008dd}), partial chemical equilibrium is not expected to play an important role. Therefore, in this paper no further discussion on partial chemical equilibrium constructions is made.

\subsection{Comparisons between S95n-v1 and the new EoS}

\begin{figure}[h]
\centering
\includegraphics[width=0.5\textwidth]{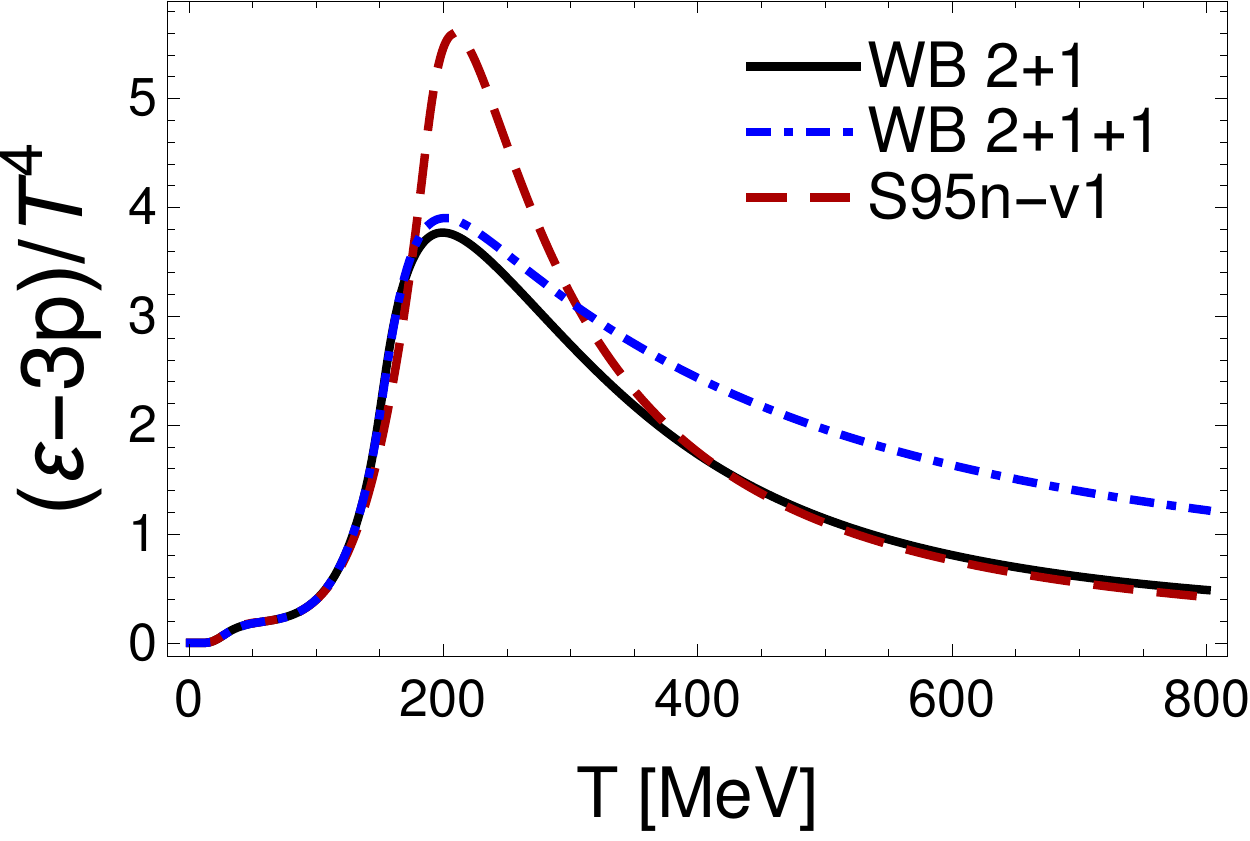} 
\caption{(Color online) Trace anomaly of the equation of state S95n-v1 \cite{Huovinen:2009yb} compared to results for the equations of state for 2+1 and 2+1+1 flavors constructed here using state-of-the-art lattice results by the Wuppertal-Budapest collaboration \cite{Borsanyi:2013bia,Borsanyi:2016ksw}.}
\label{fig:trace}
\end{figure}

\begin{figure*}[ht]
\begin{tabular}{c c}
\includegraphics[width=0.5\textwidth]{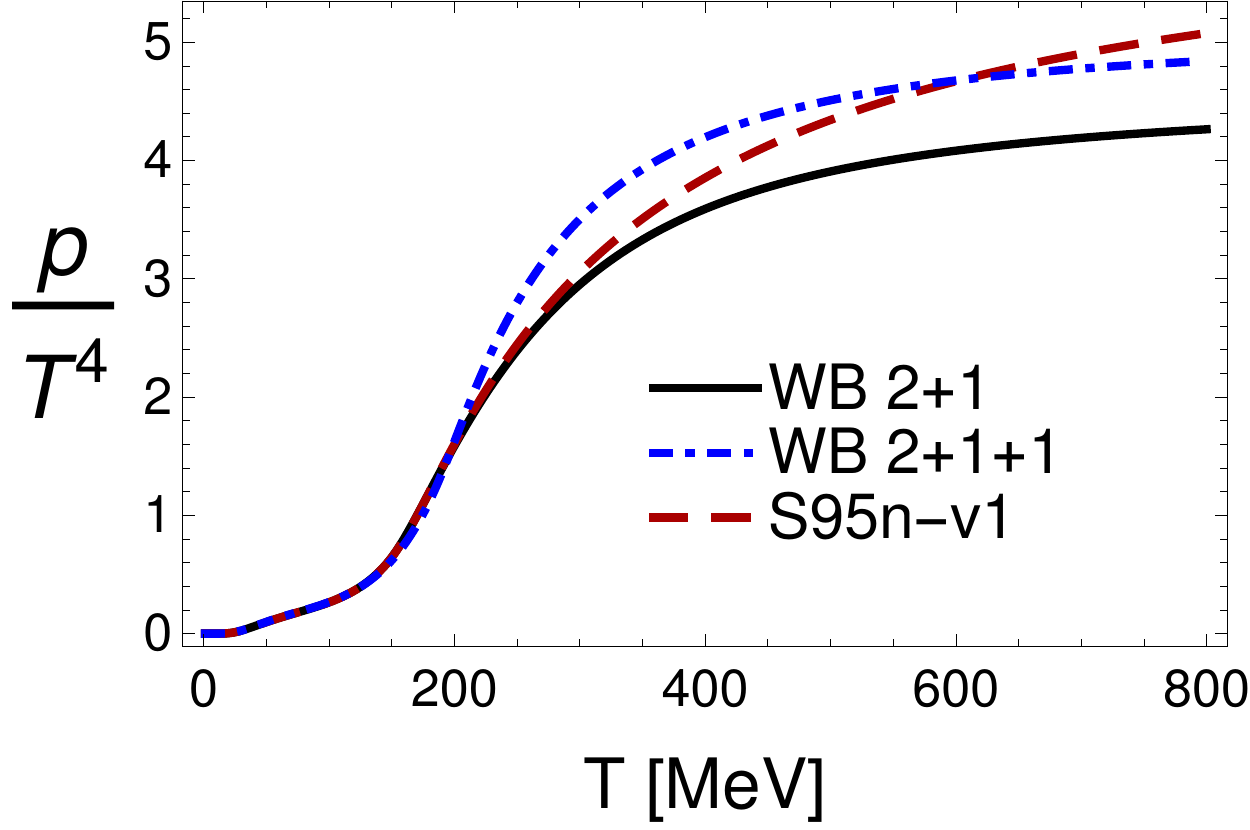}  & \includegraphics[width=0.5\textwidth]{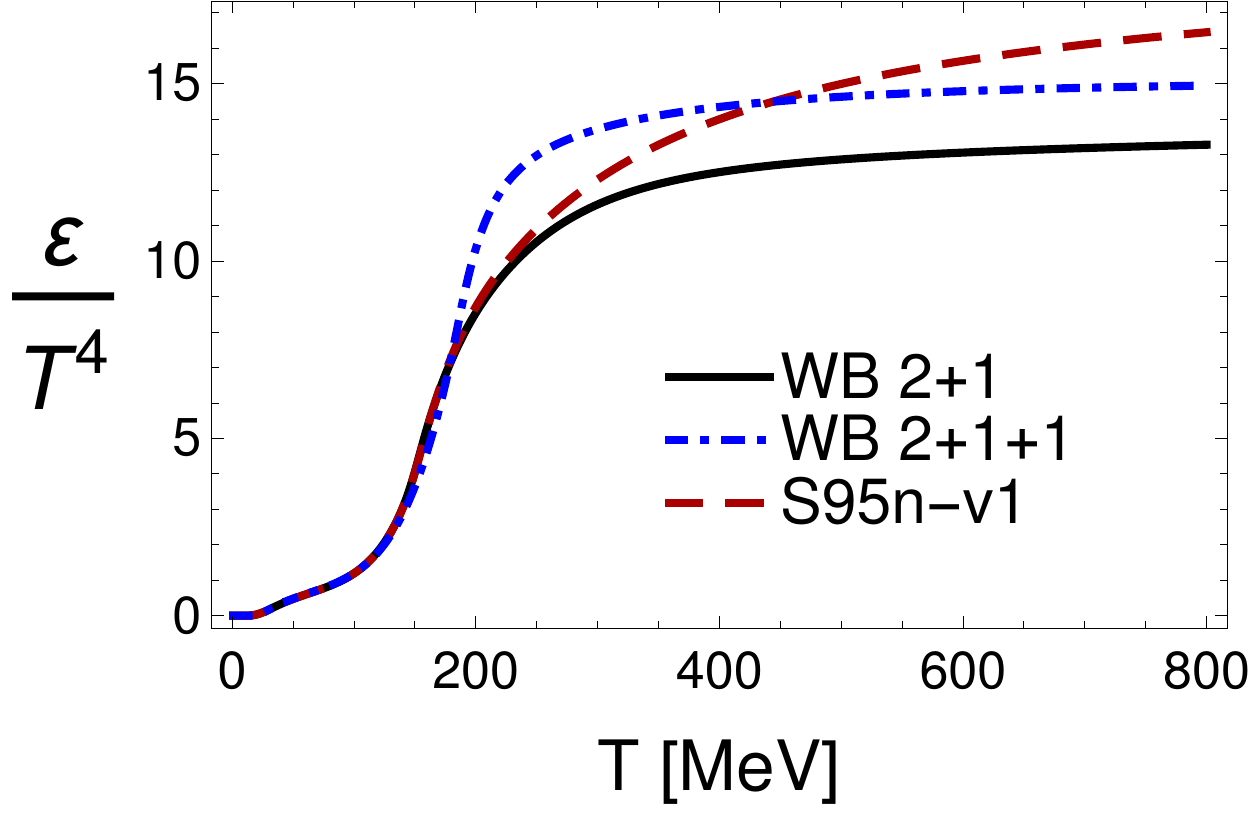} \\
\includegraphics[width=0.5\textwidth]{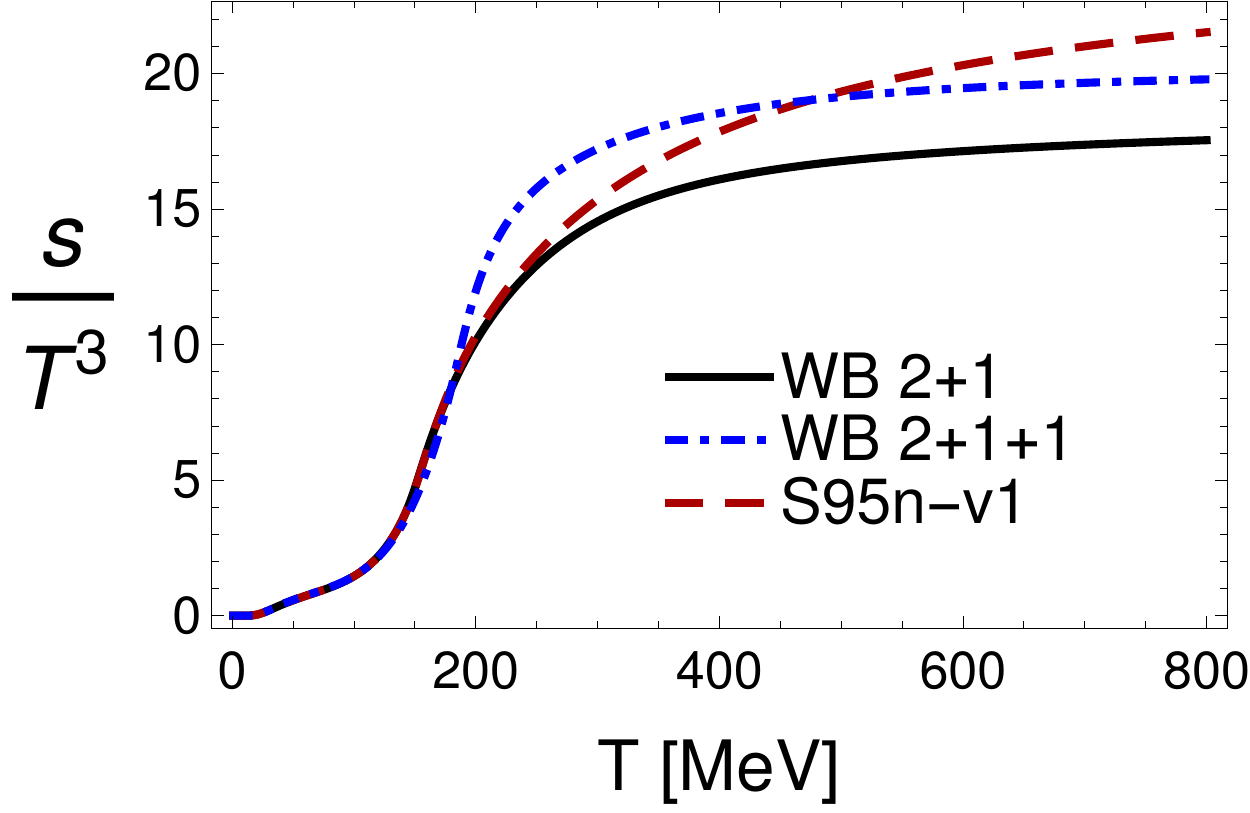}  & \includegraphics[width=0.5\textwidth]{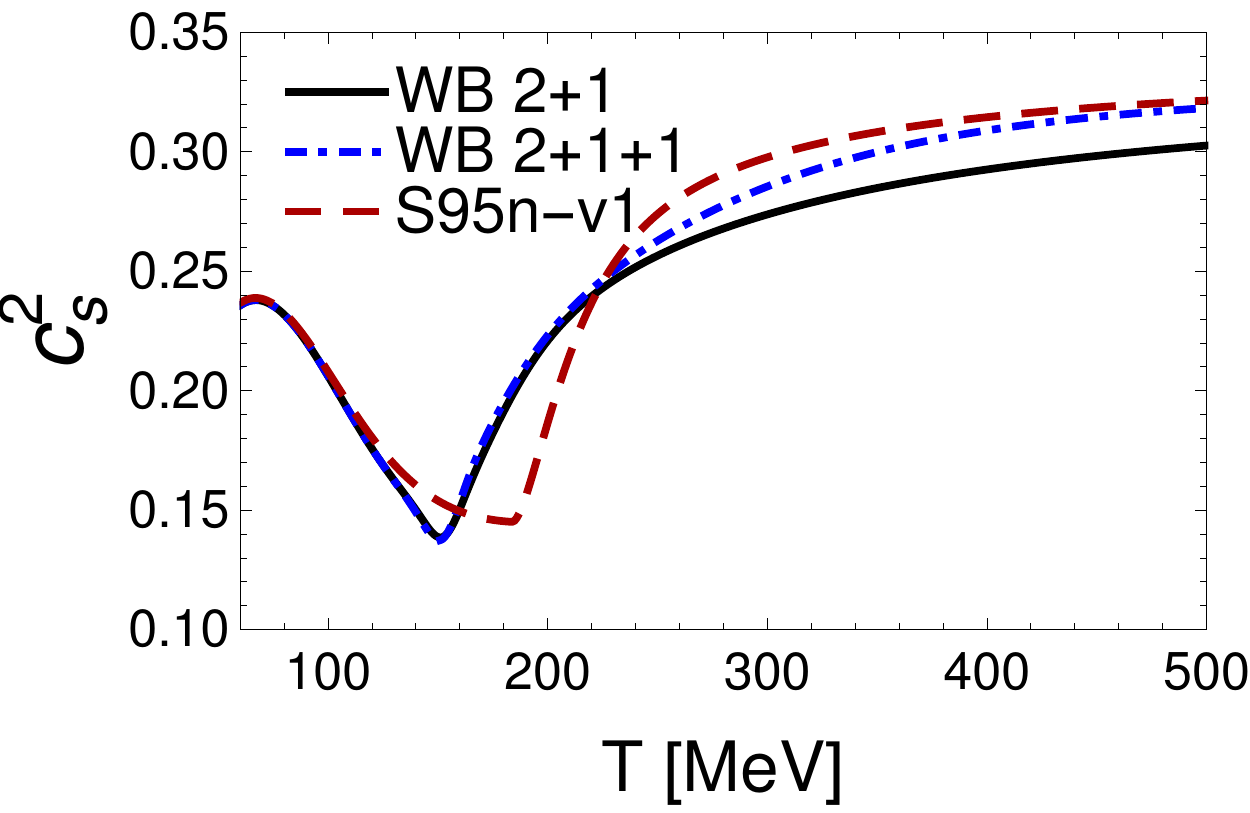} \\
\end{tabular}
\caption{(Color online) Pressure, energy density, entropy density, and speed of sound of EoS S95n-v1 \cite{Huovinen:2009yb} compared to the corresponding quantities determined using the EoS constructed here for 2+1 and 2+1+1 flavors that employed state-of-the-art lattice results by the Wuppertal-Budapest collaboration \cite{Borsanyi:2013bia,Borsanyi:2016ksw}. }
\label{fig:thermo}
\end{figure*}

The equation of state S95n-v1 presented in \cite{Huovinen:2009yb} is currently widely used in relativistic hydrodynamics and other theoretical models in the context of heavy ion collisions. In Figs.\ \ref{fig:trace}-\ref{fig:thermo} we show a comparison between S95n-v1 and the new equations of state constructed here for 2+1 and 2+1+1 flavors using state-of-the-art input from lattice and PDG\footnote{We remark that other revised equations of state have been already available for a few years, e.g., see \cite{Bluhm:2013yga}. Additionally, we note that Ref.\ \cite{Borsanyi:2016ksw} also made comparisons between different equations of state.}. 

From Fig.\ \ref{fig:trace} there is a very clear difference in the trace anomaly.  Regardless of whether charm is included or not, the new equations of state have a peak at a lower temperature and the peak is significantly lower and broader. These important differences stem from the fact that the lattice results employed in \cite{Huovinen:2009yb} were not continuum extrapolated.  As originally shown in \cite{Borsanyi:2016ksw}, the addition of thermalized charm does not change the peak of the trace anomaly but it makes the peak significantly broader by extending it to higher temperatures.  

In Fig.\ \ref{fig:thermo} we extend this comparison between equations of state and consider the other thermodynamic quantities. The most obvious difference is found in the speed of sound. Regardless of the inclusion of charm, the speed of sound of the new EoS has a minimum at a lower temperature and that minimum is also much sharper than the one found in S95n-v1, which may be relevant for the hydrodynamic evolution of the QGP. The effects of these differences in the equations of state on experimental observables will be explored in detail in Section \ref{sec:map}. 

\section{Hydrodynamical Modeling and Resonance Decays}\label{sec:hydro}

In this paper we use event-by-event fluctuating initial conditions generated by the TRENTO model \cite{Moreland:2014oya} with free parameters calibrated to fit experimental observables which have been shown to mimic the entropy deposition of saturation based calculations such as IP-Glasma \cite{Schenke:2012wb,Schenke:2014tga,Bernhard:2016tnd}. Specifically, we fix the entropy deposition parameter $p = 0$, nucleon-nucleon fluctuation shape parameter $k = 1.6$, and nucleon width $\sigma = 0.51$ motivated by fits to charged particle yields, $\langle p_T\rangle$, and event-by-event flow fluctuations \cite{Moreland:2014oya,Giacalone:2017uqx}. A very fine initial grid size of the initial conditions is set to $dx=dy=0.06$ fm at both AuAu $200$ GeV and PbPb 5.02 TeV.  First, we generate 2 million events to determine the centrality selection (based on sorting by the initial entropy).  Then, for each energy we generate 30,000 initial conditions that are run on an event-by-event basis through viscous hydrodynamics. 

We use the Smoothed Particle Hydrodynamics (SPH) Lagrangian code, v-USPhydro, to solve the viscous hydrodynamic equations taking into account shear viscous effects \cite{Noronha-Hostler:2014dqa}. The accuracy of v-USPhydro has been checked in \cite{Noronha-Hostler:2013gga,Noronha-Hostler:2014dqa} using well-known solutions of conformal hydrodynamics \cite{Marrochio:2013wla}. Viscosity is determined by fitting $v_2\{2\}$ and $v_3\{2\}$ across centrality for each equation of state individually, which will be discussed below. The three equations of state used are the PDG05/S95n-v1 \cite{Huovinen:2009yb} and PDG16+/2+1(+1)WB presented in the previous section. Hydrodynamics is switched on at $\tau_0=0.6$ fm for both RHIC and LHC run 1 and the evolution is performed using a small smoothing parameter $h=0.3$ fm (see \cite{Noronha-Hostler:2013gga,Noronha-Hostler:2014dqa,Noronha-Hostler:2015coa} for more details). At $T_{SW}=150$ MeV individual SPH particles are frozen out into hadrons \cite{Hama:2004rr} using the Cooper-Frye prescription \cite{Cooper:1974mv}. We note that for LHC run 2 using the PDG05/S95n-v1 equation of state we need to run hydrodynamics for a longer period of time to obtain reasonable results.  Thus, for this specific scenario we use $\tau_0=0.4$ fm and $T_{SW}=145$ MeV. In all the following if high statistics are needed for a specific observable we always show the effects of our sample size via jackknife resampling.

Hadronic decays are described using an adapted version of AZHYDRO \cite{Kolb:2000sd,Kolb:2002ve,Kolb:2003dz} with the full PDG16+ particle list (or PDG05 for the EoS S95n-v1). In the PDG database \cite{Patrignani:2016xqp} experimentally measured particles are classified according to their experimental certainty from * being the most uncertain states (often only marked as ``seen" without additional information) and **** being the most certain states. While thorough information is given about their mass, decay width, and isospin, information is often lacking on their decay channels into daughter particles. In the most extreme cases, * state could have no decay channels listed, for instance.  However, often a few decay channels will be listed but all of their branching ratios do not add up to $100\%$ or the decay channels will be listed without measured branching ratios. In general, these terms are intended to give a qualitative understanding of the decay properties of a particular state. Sometimes, ``seen" and ``not seen" branching ratios are listed along with other quantitative indication; ``dominant", instead, is used only in the case when no quantitative indication is present for any branching ratio but it seems rather evident that one decay mode is particularly predominant compared to the others.     

In this paper we include all listed strong decays of *-**** states with their corresponding non negligible branching ratios ($\simeq 1\%$ or higher), discarding weak decay channels. In the instances where the branching ratios do not sum to $100\%$ we assume the remaining percentage originates from radiative decays such that $N_2 \rightarrow N_1 + \gamma$ where $N_2$ and $N_1$ are hadrons with the same quantum numbers and $N_1$ is the next state in descending mass order with parity compatible for such a decay. In the cases where no quantitative information was listed, we systematically assigned a cumulative $\leq 30\%$ branching ratio to purely hadronic decays, and $\geq 70\%$ to radiative decays as explained in the previous paragraph. 

While the addition of quark model states were considered in previous comparisons to lattice QCD \cite{Bazavov:2014xya,Alba:2017mqu}, due to the further uncertainties in describing their decay channels, we leave their inclusion for a later study. Further details on the inclusion of these new resonances and their decay channels can be found in \cite{thermal}.  

We note that an important component of this paper is that when we study the addition of new resonances we alter both the equation of state and the corresponding hadronic resonances decays, unlike in previous works where the equation of state was fixed and the influence of only the decays/resonances themselves was studied \cite{Qiu:2012tm,Noronha-Hostler:2013rcw}.  We caution readers from implementing these new hadronic resonances from PDG16+ together with EoS S95n-v1 because there would not be conservation of energy as one crosses into the hadronic phase. The same could be said for using these newly created equations of state without the full feed down from  all the PDG16+ resonances.   

At this point in time no bulk viscosity is considered, which we would expect \cite{Noronha-Hostler:2013gga,Noronha-Hostler:2014dqa} to alter our $\langle p_T\rangle$ results as well as $v_n(p_T)$ \cite{Ryu:2015vwa}. Additionally, hadronic transport (such as UrQMD) is also not included here because this would require the adaptation of our particle resonance list to only include 2 body interactions as well as an adaptation of UrQMD itself to take into account the new states considered here, which is outside of the scope of this paper. Finally, only chemical equilibrium is considered and it is assumed that $\mu_B=0$, so that differences in charge are not possible and we do not anticipate a perfect fit to baryonic species. 

\subsection{Extracting $\eta/s$}\label{sec:etas}

Because first principle calculations do not yet exist for the temperature dependence of the shear viscosity, here we instead compare our theoretical results for the flow harmonics, $v_2\{2\}$ and $v_3\{2\}$, to experimental data and find a suitable constant $\eta/s$. There are a number of caveats when it comes to determining the exact temperature dependence of $\eta/s$ in the QGP from data with recent Bayesian results \cite{Bernhard:2016tnd} providing an estimate of the current uncertainty. Some of the main issues are when one switches hydrodynamics on/off, the viscous corrections to the Cooper-Frye (see, for instance, \cite{Noronha-Hostler:2013gga,Noronha-Hostler:2014dqa} for the case of v-USPhydro), possible non-flow contributions to flow harmonics, different hadronization/freeze-out temperatures for each flavor, the choice in the type of hadronic interactions following hydrodynamics, just to name a few. For simplicity's sake, we assume that the chemical freeze-out temperature is the same as the kinetic one i.e. $T_{FO}^{ch}=T_{FO}^{kin}$ \footnote{This may not be a poor assumption due to the large number of hadronic resonances considered here \cite{Broniowski:2001we}.}, which gives us smaller values of $\eta/s$ compared to other models that run hydrodynamics to lower temperatures, giving the system more time to build up flow. Furthermore, we have no initial flow/pre-equilibrium phase, which would also likely further increase $\eta/s$. Thus, we focus on the difference in the extracted $\eta/s$ due to the choice of  equation of state rather than their absolute values, as the latter will be certainly changed once some of the simplifying assumptions made here are lifted. 

\begin{table}
\begin{tabular}{ c c c}
\hline
 & AuAu 200 GeV  & PbPb 5.02 TeV \\
\hline
 PDG05/S95n-v1 & 0.05 & 0.025*\\ 
 PDG16+/2+1[WB] & 0.05 & 0.047\\  
 PDG16+/2+1+1[WB] & 0.05  &  0.04 \\
 \hline
\end{tabular}
\caption{Estimates for $\eta/s$ extracted using $v_2\{2\}$ data from RHIC and LHC. *Here it is assumed that $\tau_0=0.6$ fm and $T_{FO}=T_{kin}=150$ MeV for both RHIC and LHC with the exception of LHC for PDG05/S95n-v1 where $\tau_0=0.4$ fm and  $T_{FO}=T_{kin}=145$ MeV. Other assumptions, especially $T_{FO}> T_{kin}$ MeV, would increase $\eta/s$.}
\label{fig:etas}
\end{table}

\begin{figure*}
\centering
\begin{tabular}{cc}
\includegraphics[width=0.5\textwidth]{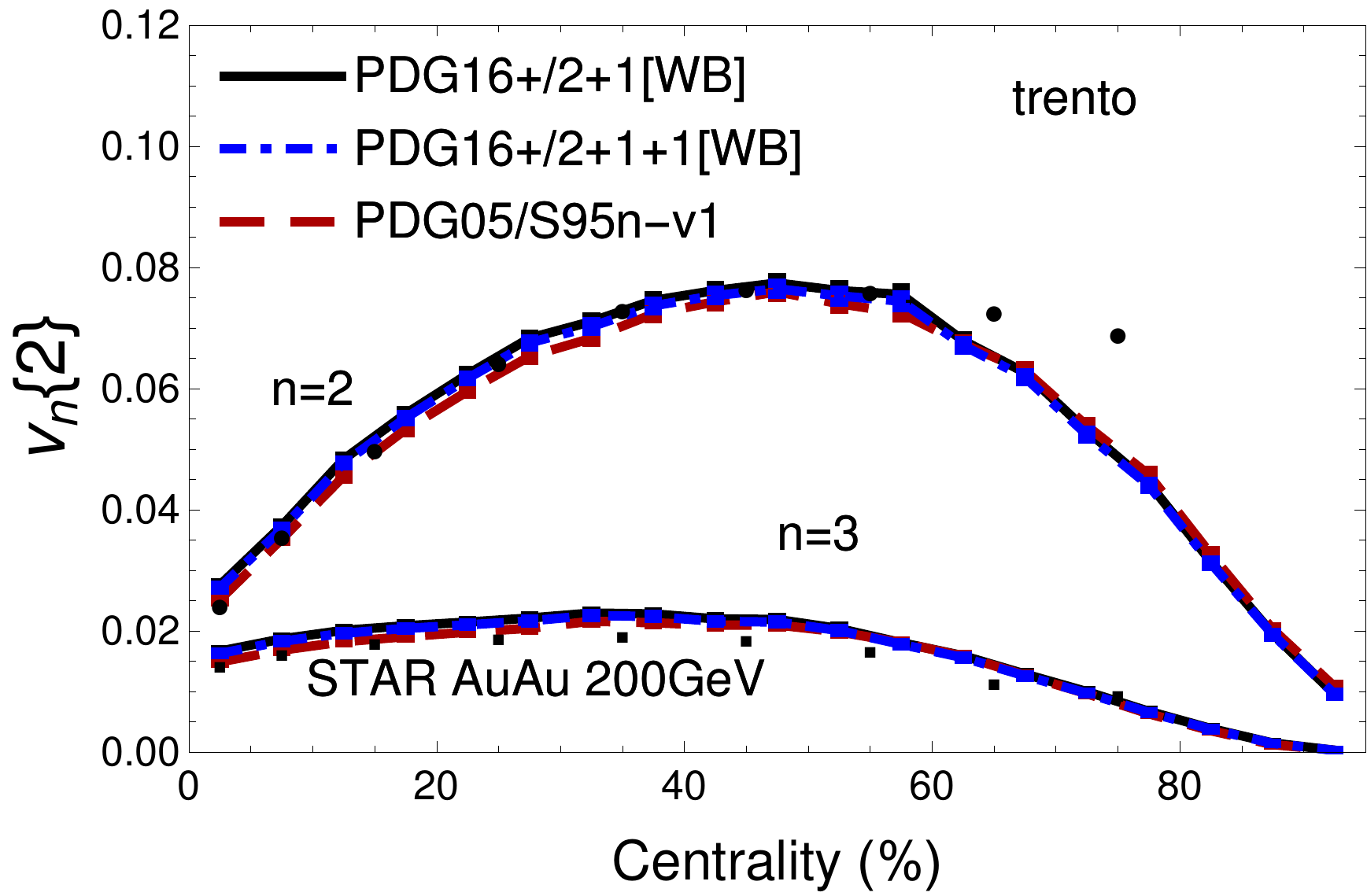} & \includegraphics[width=0.5\textwidth]{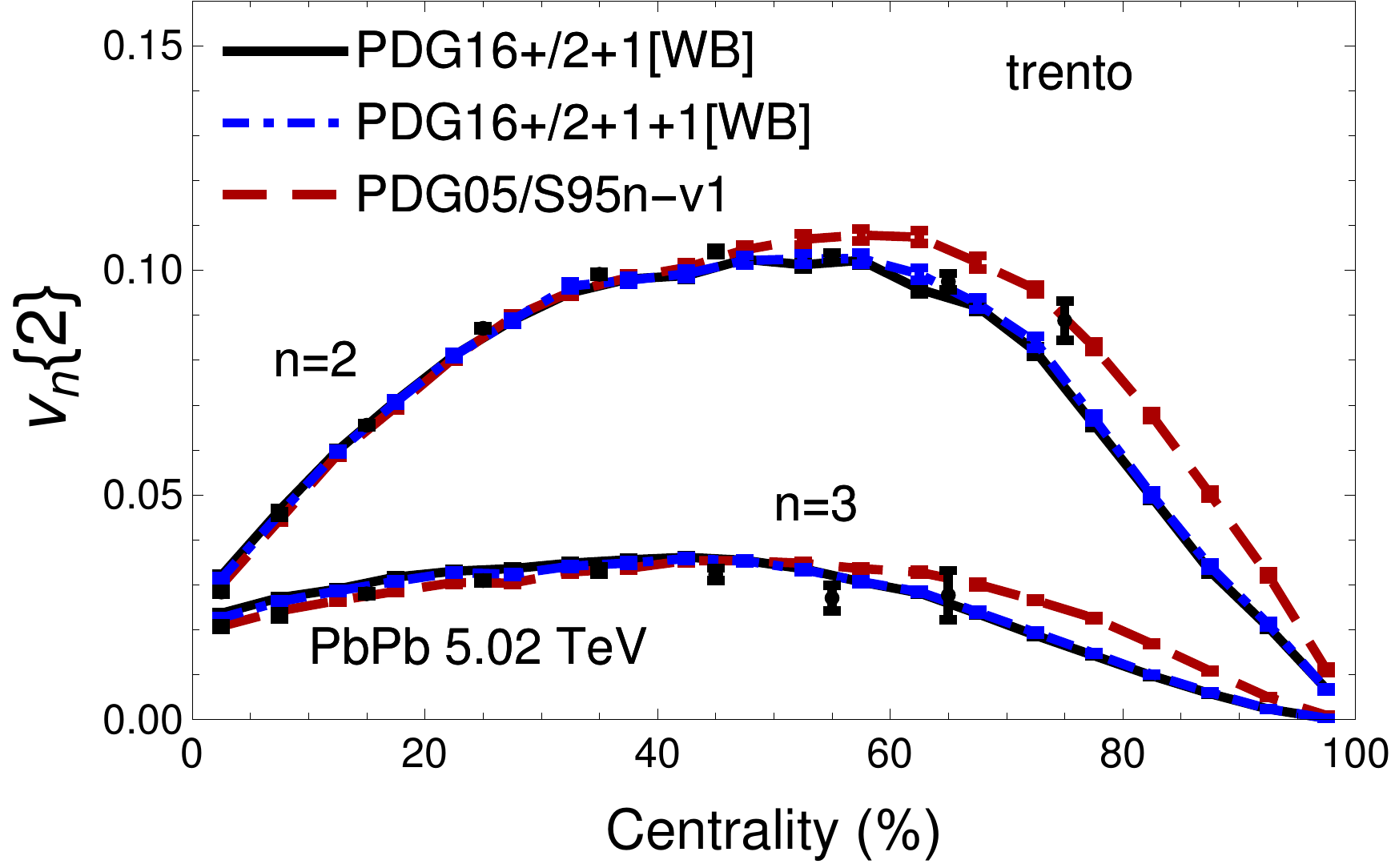} 
\end{tabular}
\caption{(Color online) $v_2\{2\}$ and $v_3\{2\}$ results across centrality for AuAu $\sqrt{s_{NN}}=200$ GeV (left) and PbPb $\sqrt{s_{NN}}=5.02$ TeV (right) collisions for all charged particles for the S95n-v1 EoS from 2009 (red, dashed lines) \cite{Huovinen:2009yb}, the 2+1 flavor WB EoS (black, full lines) \cite{Borsanyi:2013bia}, and the 2+1+1 WB EoS from 2016 (blue, dot-dashed lines) \cite{Borsanyi:2016ksw} }
\label{fig:v232}
\end{figure*}

In Table \ref{fig:etas} the extracted $\eta/s$ for each corresponding center of mass energy and EoS is shown.  Fig.\ \ref{fig:v232} shows our $v_2\{2\}$ and  $v_3\{2\}$ for each of the different equations of state compared to experimental data.  We note that for the PDG05/S95n-v1 equation of state it was very difficult to build up enough flow at LHC run 2 to match experimental data with our original parameters $\tau_0=0.6$ fm and $T_{SW}=150$ MeV and, thus, for LHC run 2 only we instead used $\tau_0=0.4$ fm and $T_{SW}=145$ MeV for that particular equation of state.  Otherwise, we used $\tau_0=0.6$ fm and $T_{SW}=150$ MeV for both RHIC and LHC run 2 to keep the parameters as similar as possible across energies.  

In Table \ref{fig:etas} one can see that at RHIC all three equations of state can use the same $\eta/s$ while still reasonably describing the flow harmonics in Fig.\ \ref{fig:v232}.  This is likely due to the fact that RHIC temperatures are quite a bit lower than LHC run 2 and most of the hydrodynamic evolution is at lower temperatures where the differences between the three equations of state are smaller. We do note, however, that the PDG05/S95n-v1 consistently produces less flow than the two more recent WB equations of state.  However, both look reasonable within error bars of the data.  At RHIC energies these differences may be a combination of the extra resonances as well as the differences between the three equations of state.

At LHC run 2, however, significant deviations for PDG05/S95n-v1 are seen in comparison to the results obtained using the WB collaboration equations of state.  Comparing  PDG05/S95n-v1 to PDG16+/WB2+1 one must increase $\eta/s$ by $88\%$ to match the experimental data.  Because no clear viscosity differences were seen at RHIC, it is safe to assume that this increase in $\eta/s$ when considering PDG16+/WB2+1 is due to the differences in the equations of state at high temperatures, not the hadronic resonances.  Comparing our results to the flow harmonics for run 2 in Fig.\ \ref{fig:v232}, we see that all three are able to match experimental data well.  However, the centrality dependence of $v_3$ differs slightly (and this may be possible to use in the future to constrain the temperature dependence of $\eta/s$). Finally, we note that there is roughly $15\%$ change between the PDG16+/WB2+1 equation of state and PDG16+/WB2+1+1, which implies that an equation of state with thermalized charmed quarks requires a slightly smaller $\eta/s$ than one with only 2+1 flavors. If we were able to probe even higher temperatures either at the LHC or a future collider then we predict an even larger splitting between the two, which has interesting implications for understanding how the shape of the equation of state relates to the build up of flow.  

In Fig.\ \ref{fig:v232} we acknowledge that we see a mismatch between our theoretical predictions and the data at very peripheral collisions.  The question remains if this is an issue with the theoretical description or could this be due to non-flow contributions in peripheral collisions?  There are strong indications that peripheral collisions are more susceptible to non-flow effects \cite{Adamczyk:2017hdl,private} so this is an interesting question for the future. We point out also that at RHIC $v_3$ is somewhat high in our calculations, which leaves room for a better fit from a temperature dependent $\eta/s$.

\subsection{Particle spectra and $\langle p_T\rangle$}
\begin{figure*}
\begin{tabular}{cc}
\includegraphics[width=0.5\textwidth]{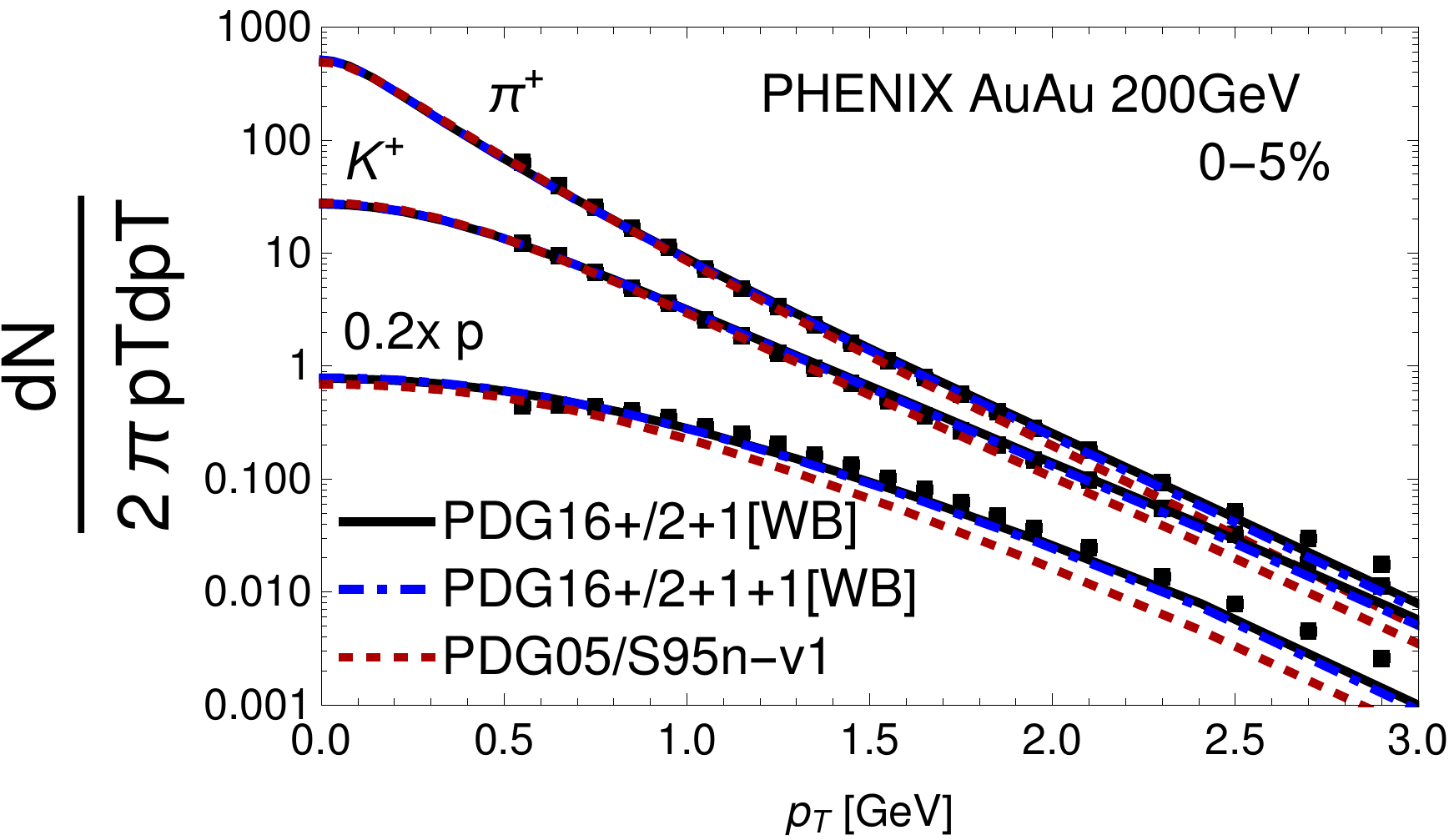} & \includegraphics[width=0.5\textwidth]{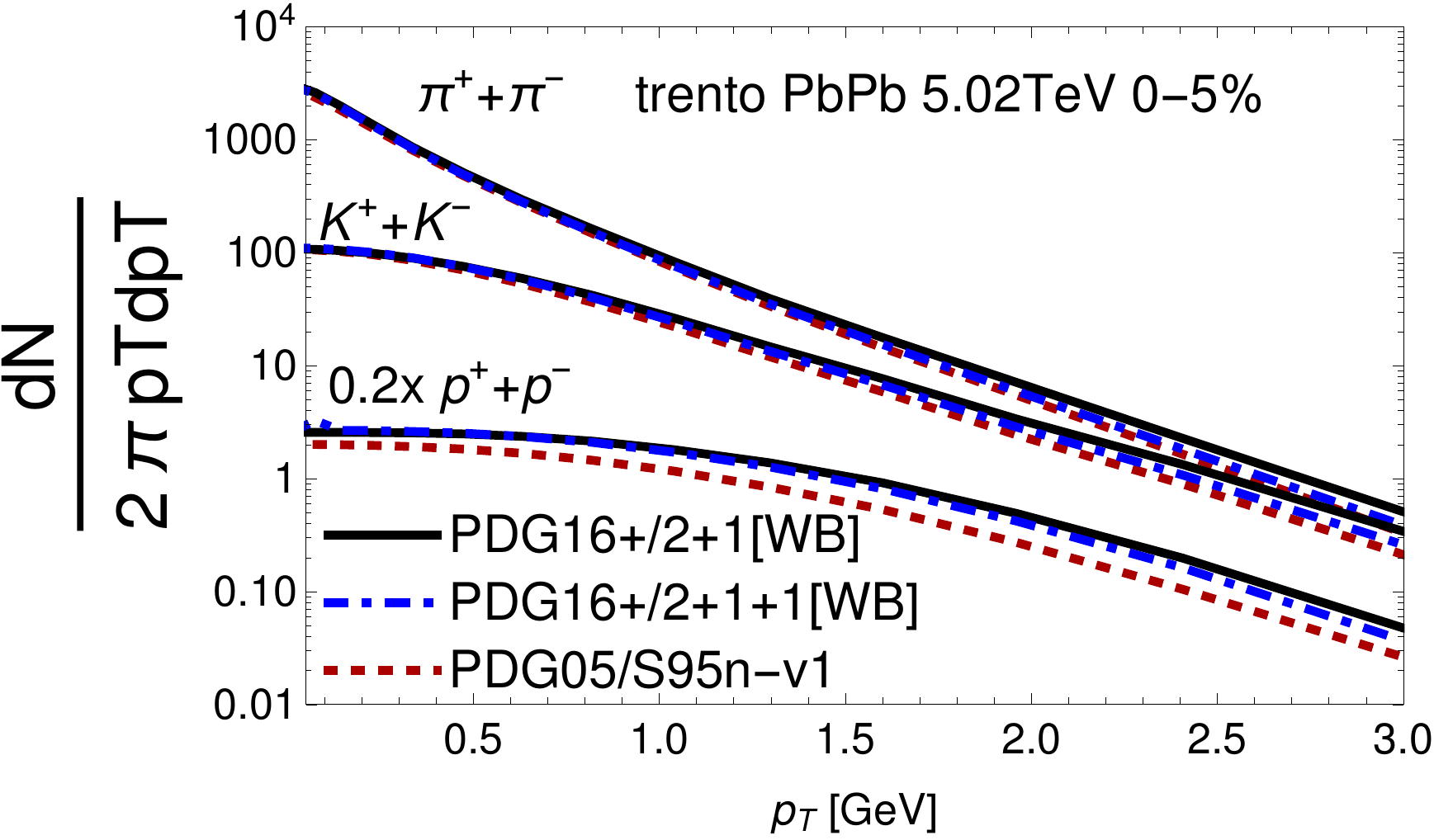} 
\end{tabular}
\caption{(Color online) Spectra of $\pi$'s, $p$'s, and $K$'s  in the centrality class $0-5\%$ for RHIC AuAu $\sqrt{s_{NN}}=200$ GeV (left) and the corresponding predictions for LHC PbPb $\sqrt{s_{NN}}=5.02$ TeV (right) collisions computed using the S95n-v1 EoS from 2009 \cite{Huovinen:2009yb}, the 2+1 WB EoS from \cite{Borsanyi:2013bia}, and the 2+1+1 WB EoS from 2016 \cite{Borsanyi:2016ksw}.  Experimental data points from PHENIX collaboration \cite{PhysRevC.69.034909} (left). }
\label{fig:pispec}
\end{figure*}

$\langle p_T\rangle$ calculations have generated a significant amount of interest in recent years due to the influence of bulk viscosity \cite{Ryu:2015vwa}.  However, one would expect that heavy resonances that decay into light particles would also affect $\langle p_T\rangle$, especially since they enhance the spectrum at high $p_T$.  Thus, here we investigate not only the effects of the three equations of state but also the influence of resonance decays on spectra and $\langle p_T \rangle$. 

Generally, we find that the biggest difference arises between PDG05/S95n-v1 vs. PDG16+/2+1(+1)WB.  The equations of state constructed using state-of-the-art lattice results produce more high $p_T$ particles and, thus, provide a better fit to experimental data as shown in Fig.\ \ref{fig:pispec} (left). This may be seen as a consequence of the sharper dip displayed by the speed of sound around the transition region found in the new EoS in comparison to the result from S95n-v1, see Fig.\ \ref{fig:thermo}.  We also show in Fig.\ \ref{fig:pispec} our predictions for the spectra at LHC run 2 (right). However, the inclusion of charm quarks into the EoS at LHC run 2 produces slightly less high $p_T$ particles in comparison to the results found using a 2+1 flavor EoS, which may also be attributed to slight differences in $\eta/s$ at LHC run 2.  

One of the biggest effects coming from the inclusion of the new hadronic resonances is the enhancement of the proton spectra and to a lesser extent the kaon spectra as well.  This enhancement occurs across all centrality classes.  We note that while our pions and kaons match experimental data well, our protons are slightly below the data.  

\begin{figure*}[ht]
\centering
\begin{tabular}{cc}
\includegraphics[width=0.5\textwidth]{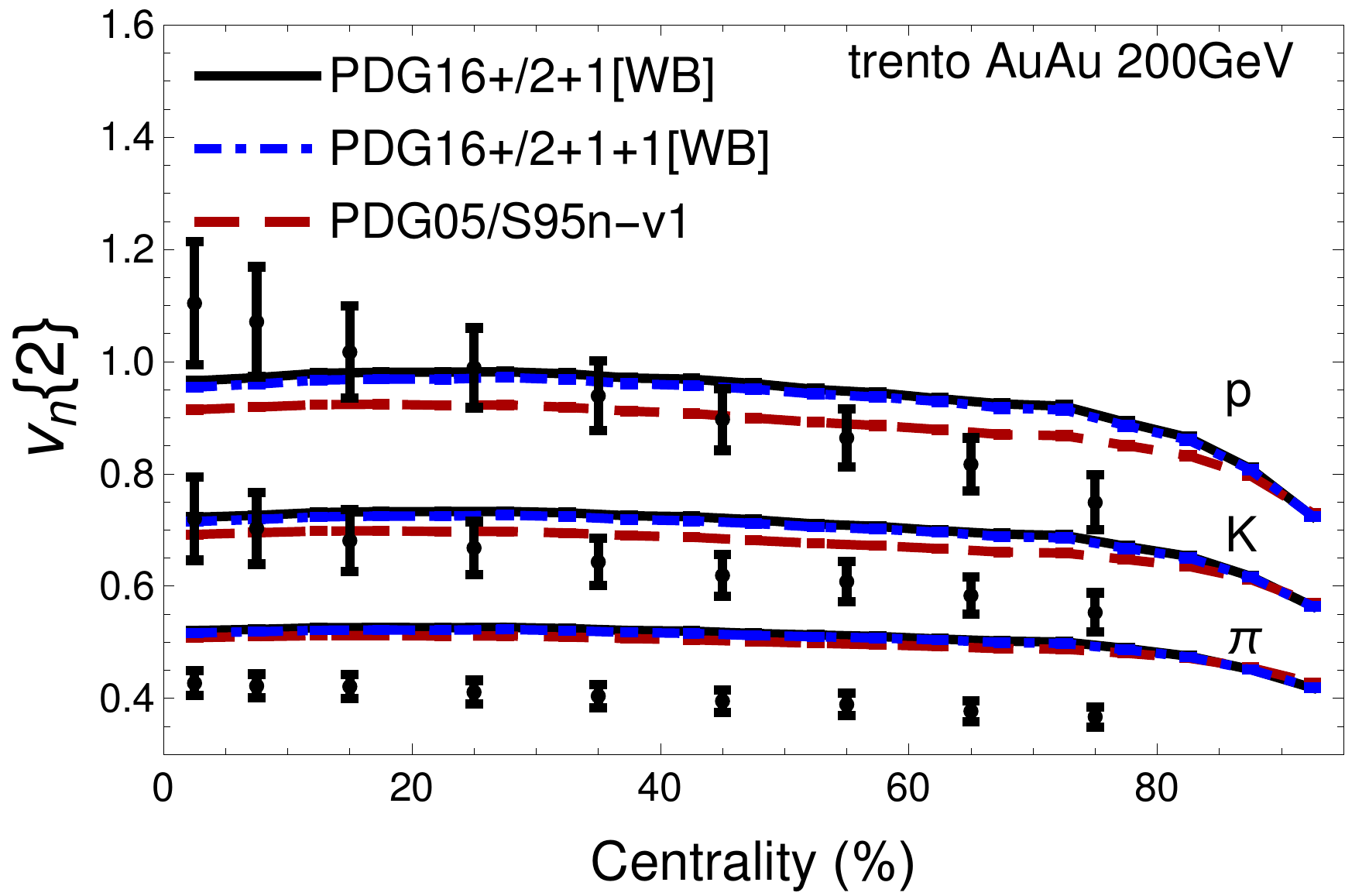} & \includegraphics[width=0.5\textwidth]{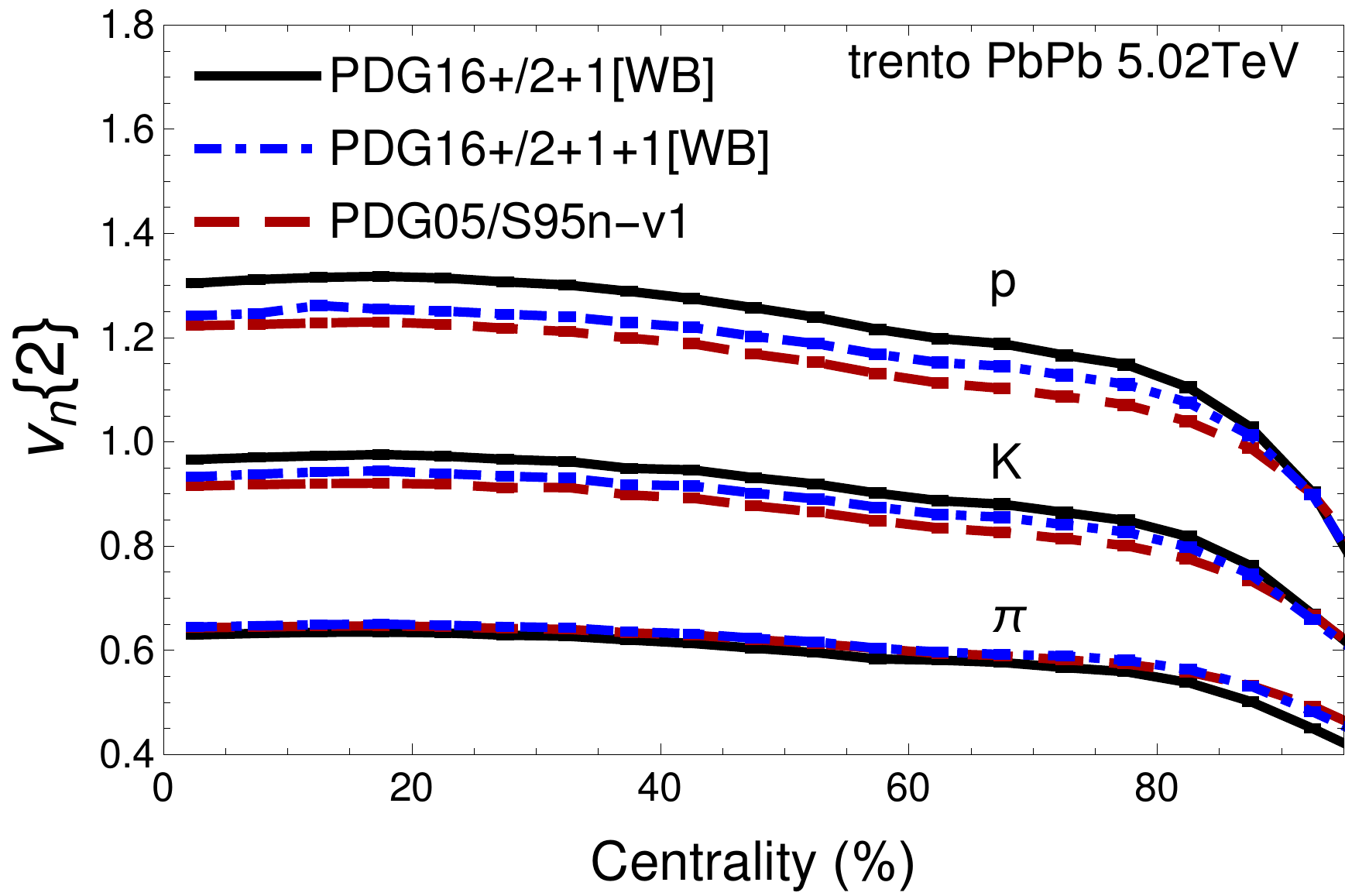} 
\end{tabular}
\caption{(Color online) $\langle p_T\rangle$ results for $\pi^+$'s, $K^+$'s and p's compared to AuAu $\sqrt{s_{NN}}=200$ GeV STAR data \cite{Abelev:2008ab} (left) and our corresponding predictions for PbPb $\sqrt{s_{NN}}=5.02$ TeV collisions (right) computed using the S95n-v1 EoS from 2009 \cite{Huovinen:2009yb}, the 2+1 WB EoS from \cite{Borsanyi:2013bia}, and the 2+1+1 WB EoS from 2016 \cite{Borsanyi:2016ksw}. }
\label{fig:meanpt}
\end{figure*}

Because we exclude the contribution from bulk viscosity we do not expect a perfect fit to $\langle p_T\rangle$ \cite{Ryu:2015vwa,Bernhard:2016tnd}. In fact, this is confirmed in Fig.\ \ref{fig:meanpt} (left) where we show our results for $\langle p_T\rangle$ across centrality for $\pi^+$'s, $K^+$'s, and p's. Our predictions for this observable at LHC run 2 are also shown in Fig.\ \ref{fig:meanpt} (right). For $K^+$'s and p's we are unable to capture the drop in $\langle p_T\rangle$ for peripheral collisions but that may also be due to a drop in the chemical/kinetic equilibrium temperatures across centralities that we do not include here \cite{Adamczyk:2017iwn}. The pion $\langle p_T\rangle$ is consistently too large regardless of the EoS.  We find that generally the extra resonances produce a larger   $\langle p_T\rangle$, as expected given our results for the spectra. Additionally, the 2+1 WB EoS gives a slightly larger  $\langle p_T\rangle$ for all hadronic species compared to the case where charm quarks are included. 

The combined effects on the particle spectra and $\langle p_T\rangle$ found here coming from using a state-of-the-art EoS and up-to-date list of resonance decays should play a role when extracting the bulk viscosity of the QGP in future calculations. Furthermore, we would expect a shift in $\langle p_T\rangle$ if these new resonances were included in a hadronic transport model. These questions are beyond the scope of this paper and are left for a future study.

\section{Results for flow correlations}\label{sec:map}

\subsection{Pearson coefficient}

It is well established that there is a strong linear correlation between the initial eccentricities and the flow harmonics, e.g., $\varepsilon_2\rightarrow v_2$ on an event-by-event basis \cite{Teaney:2010vd,Gardim:2011xv,Niemi:2012aj,Teaney:2012ke,Qiu:2011iv,Gardim:2014tya,Betz:2016ayq}.  One method of quantifying this is using a Pearson coefficient \cite{Gardim:2011xv,Gardim:2014tya,Betz:2016ayq} involving the flow vectors $\left\{v_n,\psi_n\right\}$ and the eccentricities $\left\{\varepsilon_n,\phi_n\right\}$ such that:
\begin{equation}
Q_n=\frac{\langle v_n \varepsilon_n \cos\left(n\left[\psi_n-\phi_n\right]\right)\rangle}{\sqrt{\langle |\varepsilon_n|^2\rangle \langle  |v_n|^2 \rangle}}
\end{equation}
where a value of $Q_n=1$ indicates a perfect linear correlation between the initial eccentricities and the final flow harmonics whereas 0 means there is no linear correlation.  We emphasize that this definition of the Pearson coefficient considers both the magnitude and the angle of the flow harmonics so this implies that the entire eccentricity {\it vector} is correlated with the final flow harmonic {\it vector}.  

\begin{figure*}[ht]
\centering
\begin{tabular}{cc}
\includegraphics[width=0.5\textwidth]{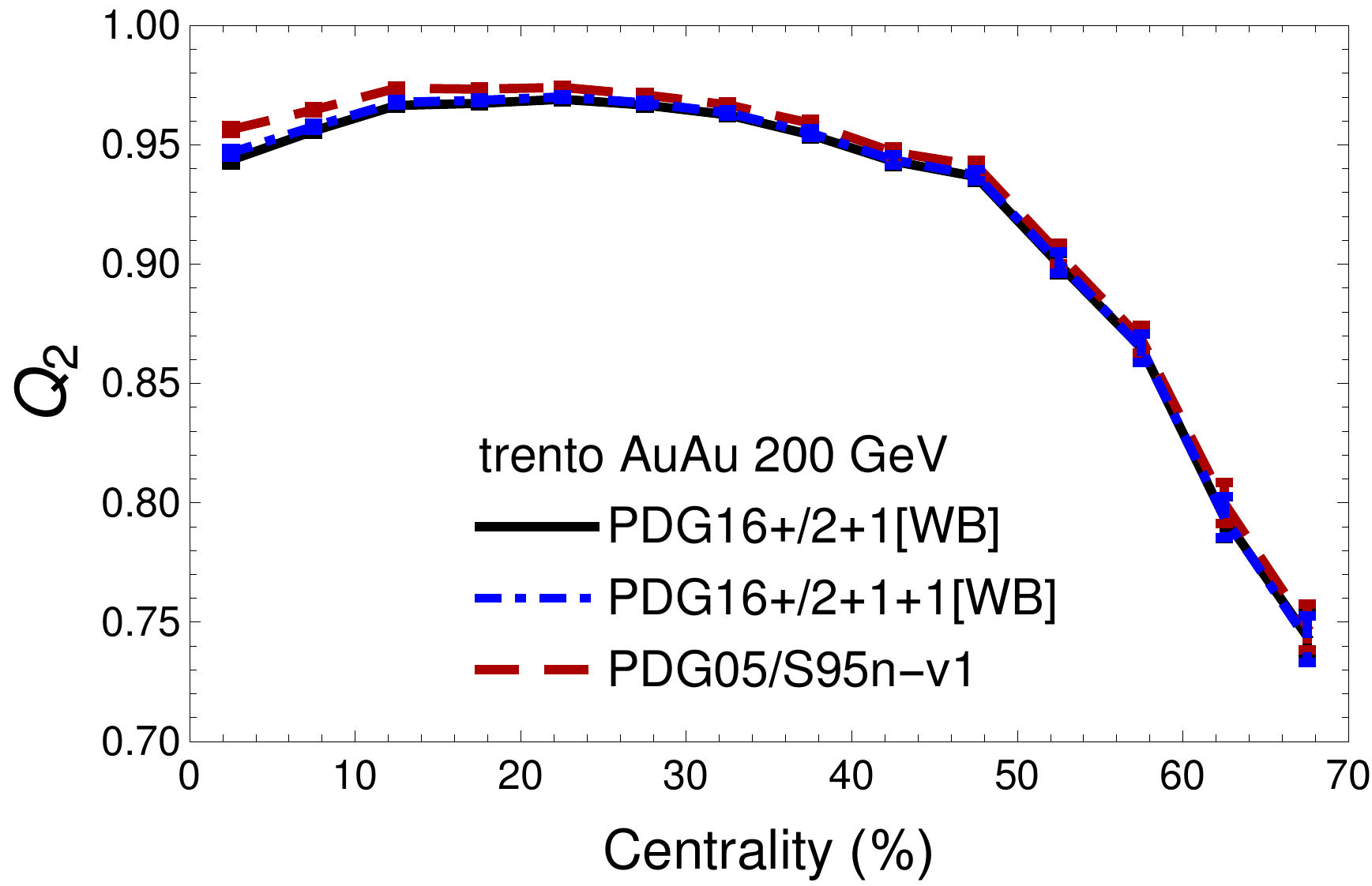} & \includegraphics[width=0.5\textwidth]{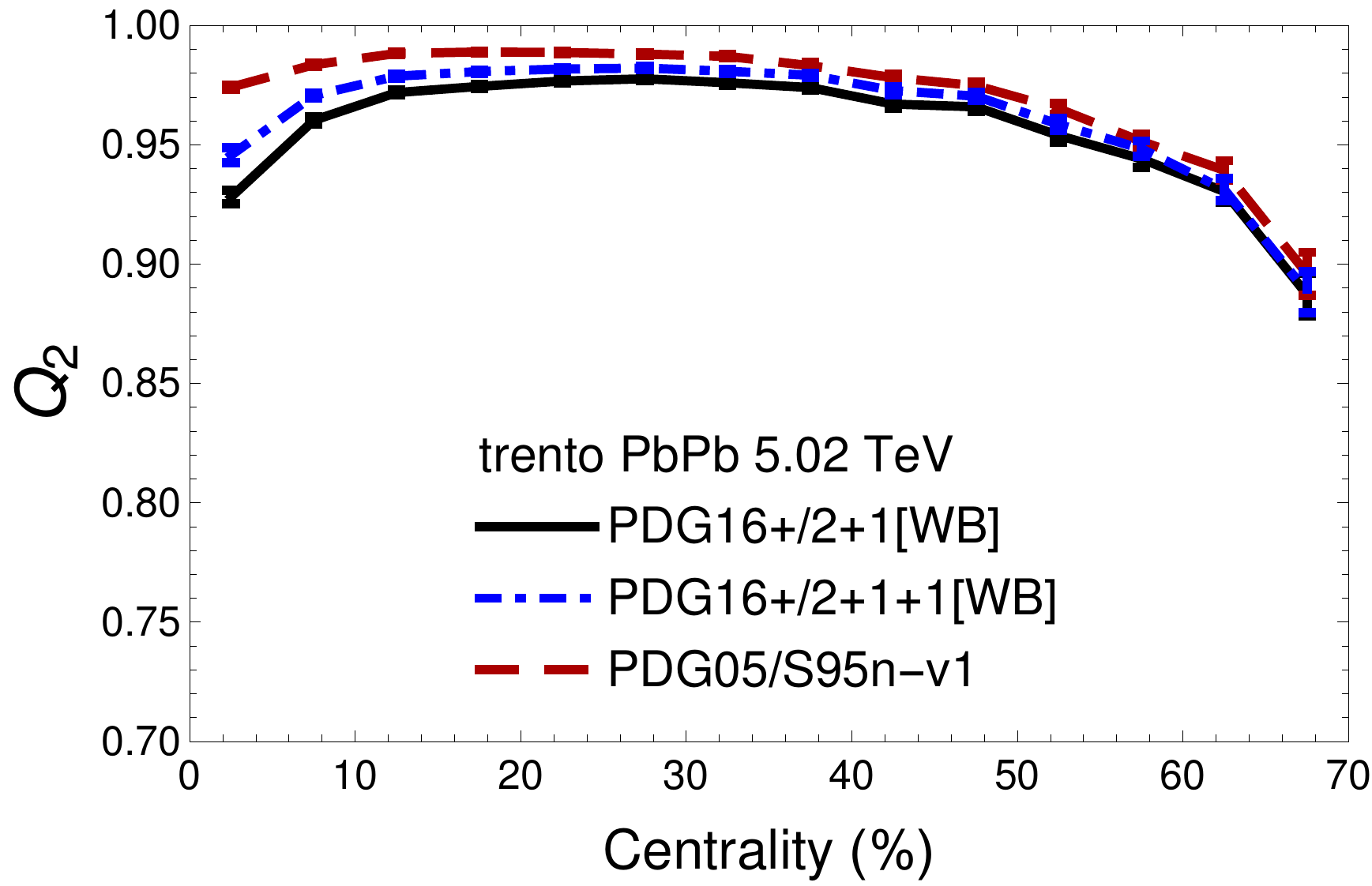} 
\end{tabular}
\caption{(Color online) Pearson coefficient results for $v_2$ of all charged particles 
for AuAu $\sqrt{s_{NN}}=200$ GeV collisions (left) and PbPb $\sqrt{s_{NN}}=5.02$ TeV collisions (right) computed using the S95n-v1 EoS from 2009 \cite{Huovinen:2009yb}, the 2+1 WB EoS from \cite{Borsanyi:2013bia}, and the 2+1+1 WB EoS from 2016 \cite{Borsanyi:2016ksw}.}
\label{fig:q2pions}
\end{figure*}
In Fig.\ \ref{fig:q2pions} (left) the Pearson coefficient corresponding to the mapping $\varepsilon_2\rightarrow v_2$ of all charged hadrons is shown and one can see that the choice of EoS essentially has no influence at RHIC except perhaps for very central collisions. In Fig.\ \ref{fig:q2pions} (right) the corresponding calculation at LHC run 2 is shown. In this case one can see larger effects due to the choice of the EoS, especially in more central collisions. The Pearson coefficient for PDG05/S95n-v1 is closer to unity being larger than the one found using the other equations of state constructed using state-of-the-art lattice results. Overall, the linear mapping between $v_2$ and $\varepsilon_2$ is still very good. Fig.\ \ref{fig:q2pions} also shows how this mapping changes with $\sqrt{s_{NN}}$. At the highest LHC energies there is a strong linear mapping all the way to peripheral collisions ($Q_2 \gtrsim 0.9$) whereas for RHIC $\sqrt{s_{NN}}=200$ GeV the Pearson coefficient drops more significantly in peripheral collisions, which indicates that in this regime non-linear contributions have become relevant. This may be a consequence of the shorter lifetime (smaller volume size) of the QGP formed at RHIC vs. LHC run 2. 

\begin{figure}[ht]
\centering
\includegraphics[width=0.5\textwidth]{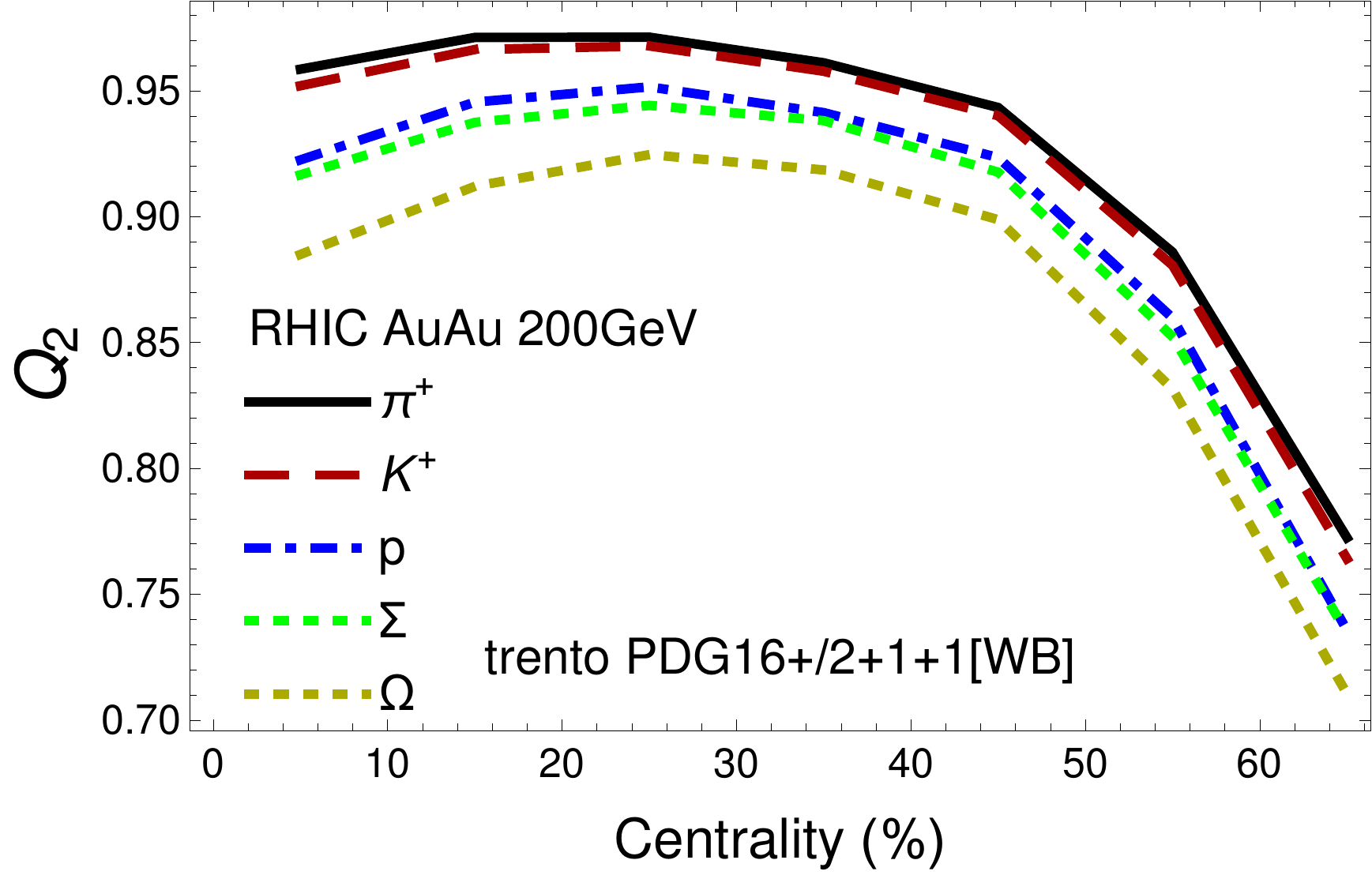} 
\caption{(Color online) Pearson coefficient results for $v_2$ of various identified particles in AuAu $\sqrt{s_{NN}}=200$ GeV collisions (left) and PbPb $\sqrt{s_{NN}}=5.02$ TeV collisions (right) computed using the 2+1+1 WB EoS \cite{Borsanyi:2016ksw}.}
\label{fig:q2pid}
\end{figure}
While it is now clear that there is a strong linear correlation between the initial eccentricity and the elliptic flow of all charged particles, one may wonder how this correlation changes for identified hadrons. In Fig.\ \ref{fig:q2pid} we present the Pearson coefficient for identified particles, which shows that the elliptic flow of heavier particles is less linearly correlated to the initial eccentricity in comparison to result found for light particles.  This suggests that other non-linear effects \cite{Noronha-Hostler:2015dbi} may play a more relevant role for heavier hadrons. In fact, if non-linear contributions to flow possess a mass dependence this could be later explored to investigate medium effects such as viscosity. Questions still remain regarding differences in the light vs. strange chemical equilibrium temperatures \cite{Bellwied:2013cta,Noronha-Hostler:2016rpd} so it is not yet clear how the results in Fig.\ \ref{fig:q2pid} would be affected if strange hadrons were formed earlier in the hydrodynamic evolution. If $T_{strange}^{ch}>T_{light}^{ch}$, it may be that the larger deviation from unity found for the Pearson coefficients of strange hadrons in Fig.\ \ref{fig:q2pid} could be further enhanced. We leave this for a future study.

\subsection{$v_2\{2\}/v_3\{2\}$ puzzle in ultracentral collisions}

Generally, in non-central collisions one expects that there is a clear hierarchy in the flow harmonics driven by both geometric effects and also viscosity, i.e., $v_2\{2\}>v_3\{2\}>v_4\{2\}$ etc. However, in ultracentral collisions all the eccentricities are fluctuation-driven and therefore equivalent, i.e., $\varepsilon_2 \sim \varepsilon_3$, which would imply, e.g., that $v_2\{2\} > v_3\{2\}$ since in hydrodynamic calculations higher harmonics are more suppressed by viscosity. 

The inability of model calculations to describe the surprising result that $v_2\{2\}\sim v_3\{3\}$ in ultracentral collisions \cite{CMS:2013bza} remains a major puzzle in the field. In fact, this result has not yet been explained by hydrodynamical models \cite{Shen:2015qta} though it has been suggested that bulk viscosity could a play role in its explanation \cite{Rose:2014fba}. Additionally, it is not only the two particle cumulant that has issues in ultracentral collisions since $v_3$ fluctuations also underpredict experimental data \cite{Giacalone:2017uqx}. 

\begin{figure*}[ht]
\centering
\begin{tabular}{cc}
\includegraphics[width=0.5\textwidth]{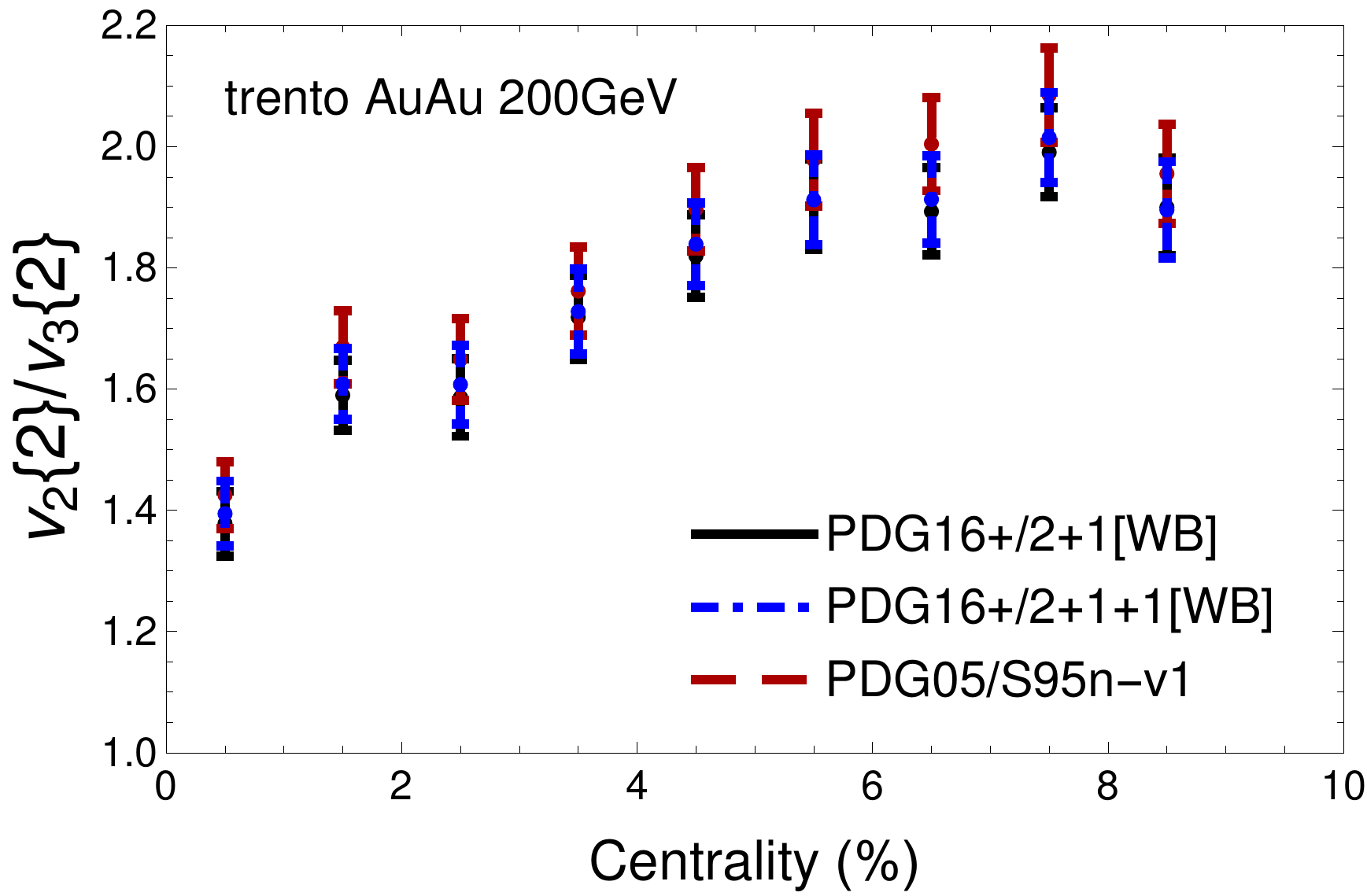} & \includegraphics[width=0.5\textwidth]{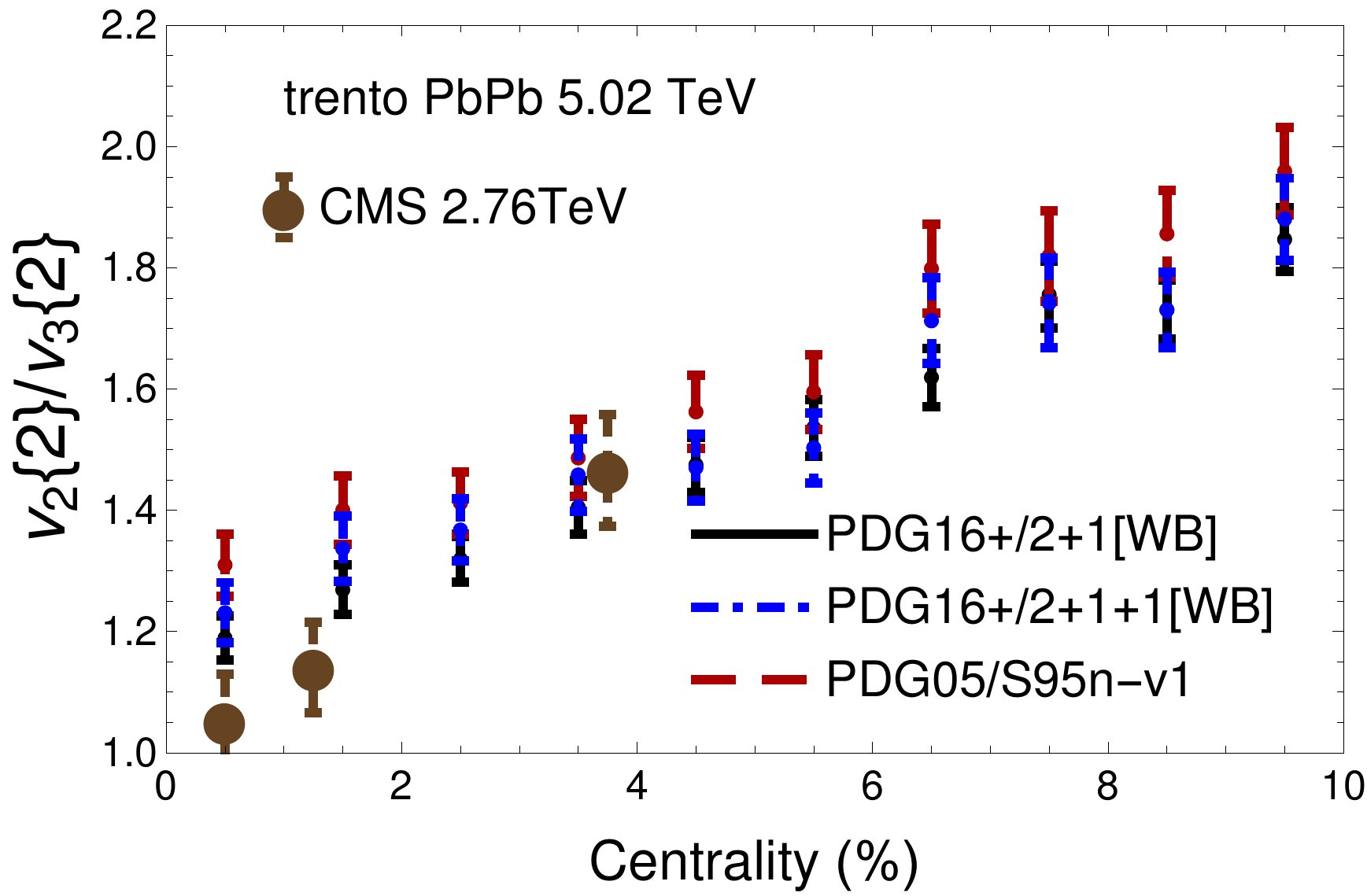} 
\end{tabular}
\caption{(Color online) $v_2\{2\}/v_3\{2\}$ results in AuAu $\sqrt{s_{NN}}=200$ GeV collisions (left) and PbPb $\sqrt{s_{NN}}=5.02$ TeV collisions (right) for all charged particles computed using the S95n-v1 EoS from 2009 \cite{Huovinen:2009yb}, the 2+1 WB EoS from \cite{Borsanyi:2013bia}, and the 2+1+1 WB EoS from 2016 \cite{Borsanyi:2016ksw}.  Experimental data for LHC $\sqrt{s_{NN}}=2.76$ TeV from CMS \cite{CMS:2013bza} are included for comparison (error propagation was implemented to obtain the error of the ratio, which was not originally presented by CMS).}
\label{fig:v23}
\end{figure*}

In this paper we checked if the choice of the equation of state could affect the ratio $v_2\{2\}/v_3\{2\}$ in ultracentral collisions. Our results for RHIC (left) and LHC run 2 (right) are shown in Fig.\ \ref{fig:v23}. We note that this ratio was not provided by CMS so we used simple error propagation to obtain the corresponding error bar. Additionally, experimental data is not yet available for LHC run 2 for ultracentral collisions so we used $\sqrt{s_{NN}}=2.76$ TeV data for  comparison.

In Fig.\ \ref{fig:v23} one can immediately see that hydrodynamic calculations of the ratio $v_2\{2\}/v_3\{2\}$ have a clear beam energy dependence.  At RHIC for $0-1\%$ centrality our calculations give $v_2\{2\}/v_3\{2\}\sim 1.4$ whereas at LHC run 2 we find a lower value $v_2\{2\}/v_3\{2\}\sim 1.2$ (depending on the equation of state). We note that the improvement in the equation of state made here did not solve this puzzle in ultracentral collisions. However, it would be interesting to see if our hydrodynamic model is able to describe at least the general trend, i.e., the prediction made here that this ratio increases when going to lower $\sqrt{s_{NN}}$. This could be verified via a measurement of $v_2\{2\}/v_3\{2\}$ also in ultracentral collisions at RHIC energies. One possible explanation for the small value of this ratio at LHC is that $v_2\{2\}$ saturates to experimental values sooner in the evolution than $v_3\{2\}$, which would then imply that the typical hydrodynamic evolution used in simulations is not yet long enough for ultracentral collisions at the LHC.  However, with our current results we cannot make a conclusive statement.  

One final point to be made is that $v_2\{2\}/v_3\{2\}$ does appear to have some dependence on the equation of state. However, at this point we do not have enough statistics to make a conclusive statement.  Nevertheless, it appears that there is a slight improvement when going from PDG05/S95n-v1 to  PDG16+/2+1(+1)WB at LHC run 2. It is also quite possible that this could be a centrality binning issue due to experimental error for large multiplicities.  For instance, in the next section it is shown that in central collisions the fluctuations are quite large, which implies that any error in the centrality binning could cause errors in the quickly changing difference between $v_2\{2\}$ and $v_3\{2\}$ in that regime.


\subsection{Flow fluctuations}

Since the discovery of triangular flow and the advent of event-by-event hydrodynamic simulations (for a review see \cite{Luzum:2013yya}), a large body of research has been developed on the subject of flow fluctuations, e.g., \cite{Borghini:2001vi,Bilandzic:2010jr,Bhalerao:2014xra,Noronha-Hostler:2015dbi,Giacalone:2016mdr,Betz:2016ayq,Giacalone:2016eyu,Giacalone:2017uqx}.  Within a set centrality class a wide distribution of flow harmonics are measured by ATLAS \cite{Aad:2013xma} wherein one can then describe the moments of the distribution via multiparticle cumulants:
\begin{eqnarray}\label{eqn:cumulants}
\nonumber v_n\{2\}^2 &=& \langle v_n^2 \rangle, \\
\nonumber v_n\{4\}^4 &=& 2 \langle v_n^2\rangle^2 - \langle v_n^4\rangle, \\
\nonumber v_n\{6\}^6 &=& \frac{1}{4} \biggl [ \langle v_n^6 \rangle - 9 \langle v_n^2\rangle \langle v_n^4 \rangle + 12 \langle v_n^2\rangle^3 \biggr], \\
\nonumber v_n\{8\}^8 &=& \frac{1}{33} \biggl [ 144 \langle v_n^2\rangle^4 - 144 \langle v_n^2\rangle^2 \langle v_n^4 \rangle + 18 \langle v_n^4\rangle^2 \\
 &+& 16 \langle v_n^2 \rangle \langle v_n^6 \rangle - \langle v_n^8 \rangle \biggr ],\nonumber
\end{eqnarray}
where $\langle v_n^2 \rangle$ is averaged over events within a set centrality class.  Normally, multiplicity weighing and centrality rebinning is used experimentally \cite{Bilandzic:2010jr,Bilandzic:2013kga}, which do have some effects in central and peripheral collisions in theoretical calculations as well \cite{Gardim:2016nrr,Betz:2016ayq}.  Here, however, we do not use them due to statistical limitations but we may explore this option in a future work with higher statistics.  

\begin{figure*}
\centering
\begin{tabular}{cc}
\includegraphics[width=0.5\textwidth]{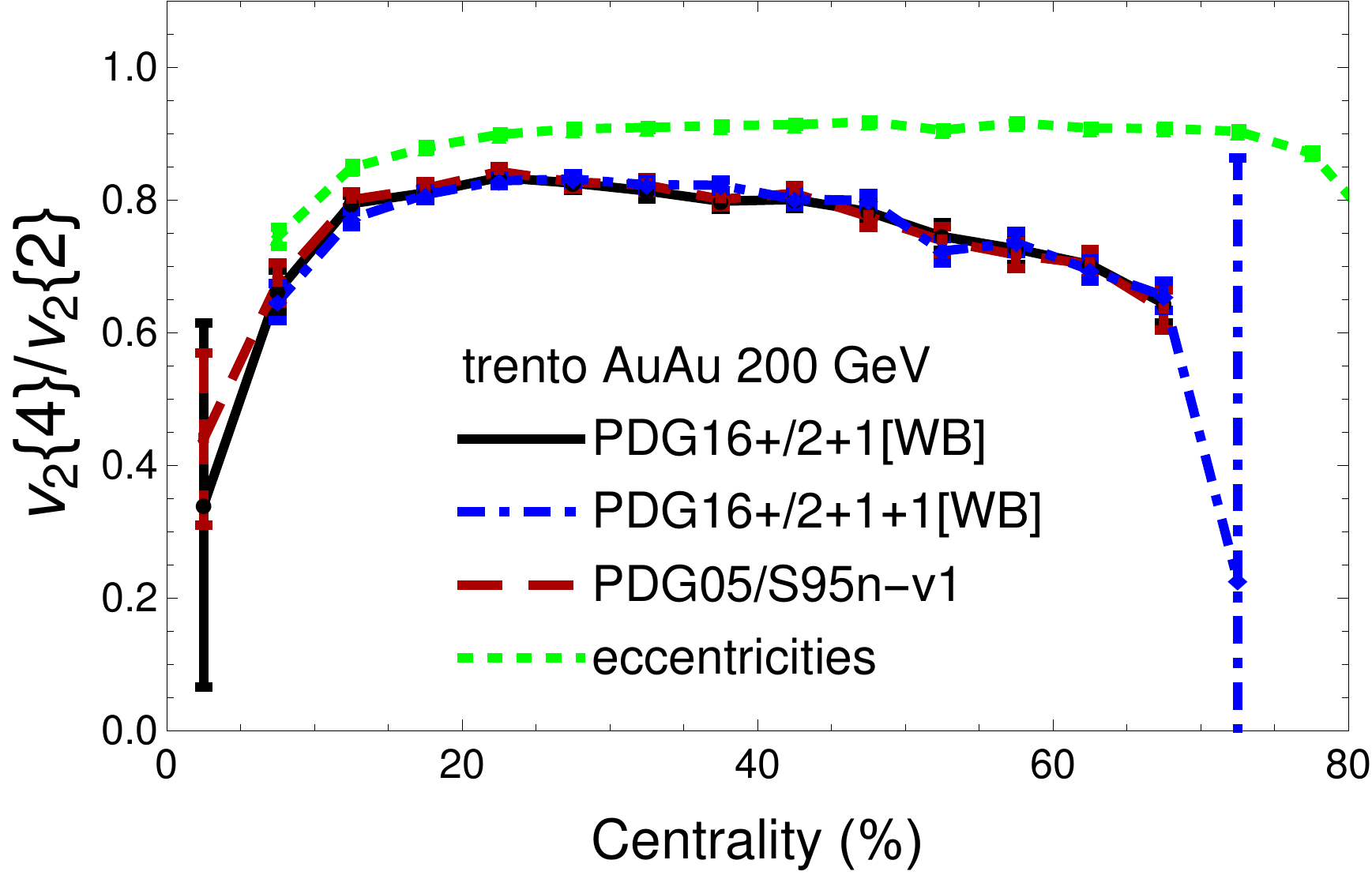} & \includegraphics[width=0.5\textwidth]{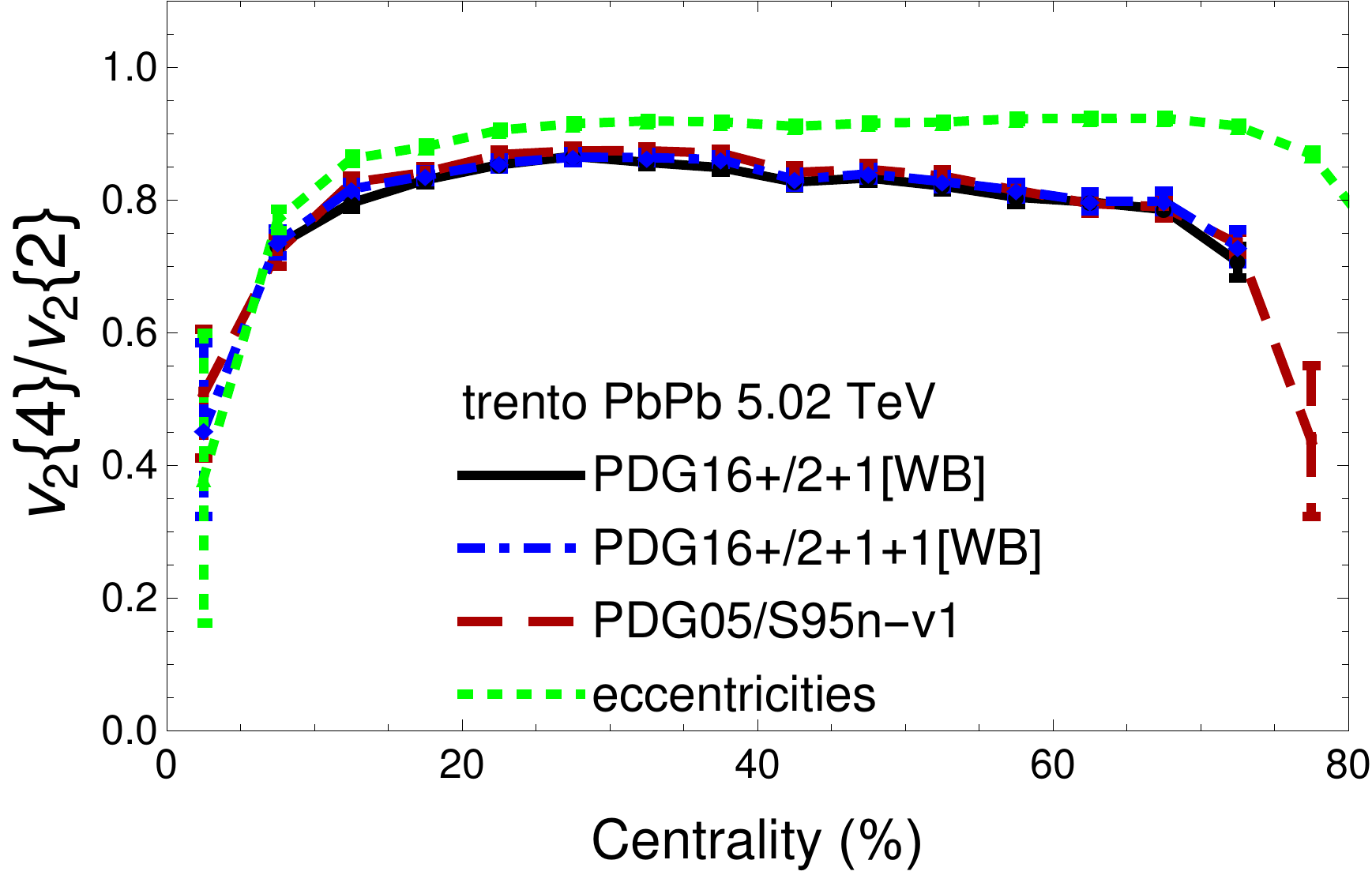} \\
\includegraphics[width=0.5\textwidth]{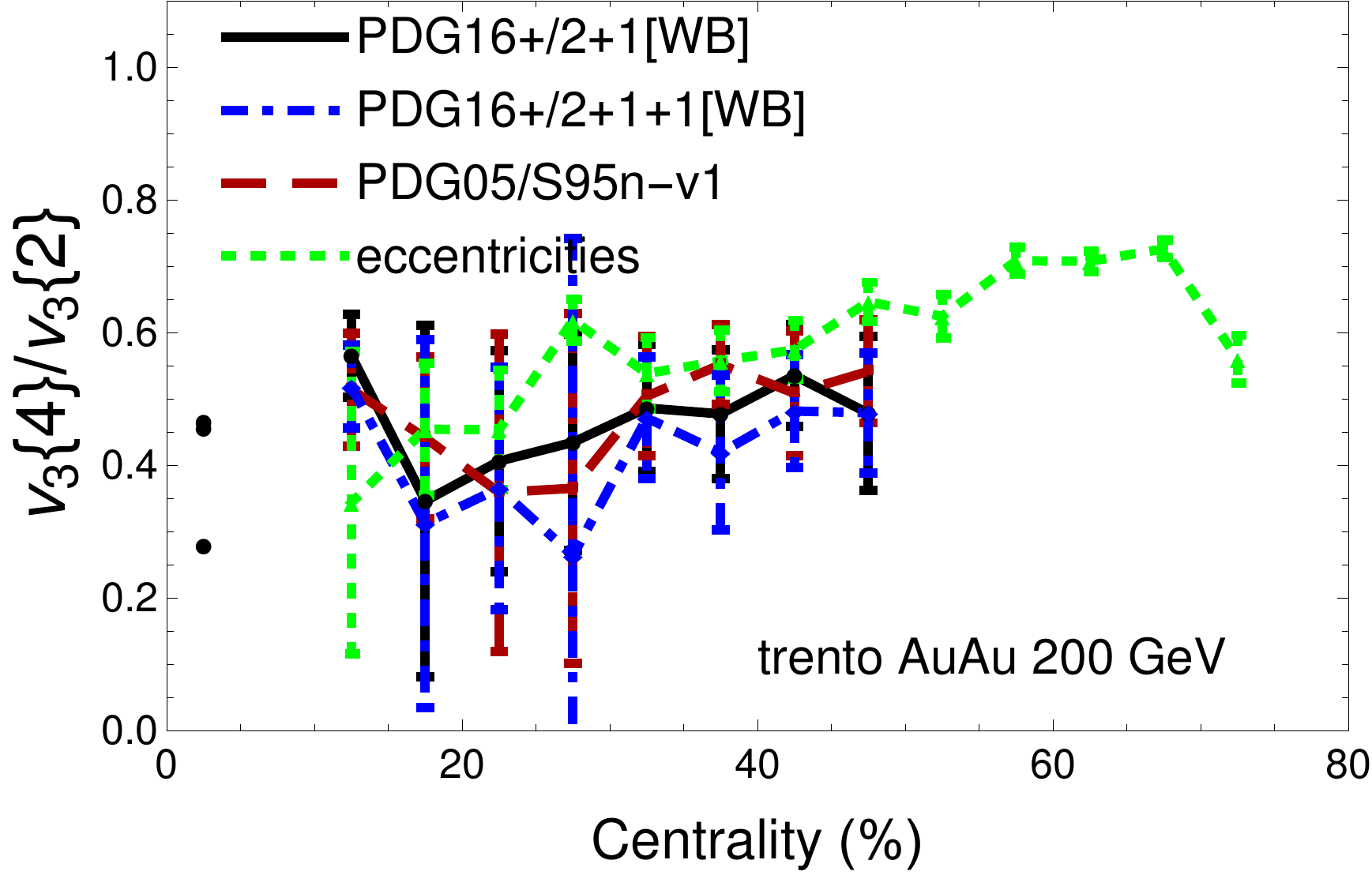} & \includegraphics[width=0.5\textwidth]{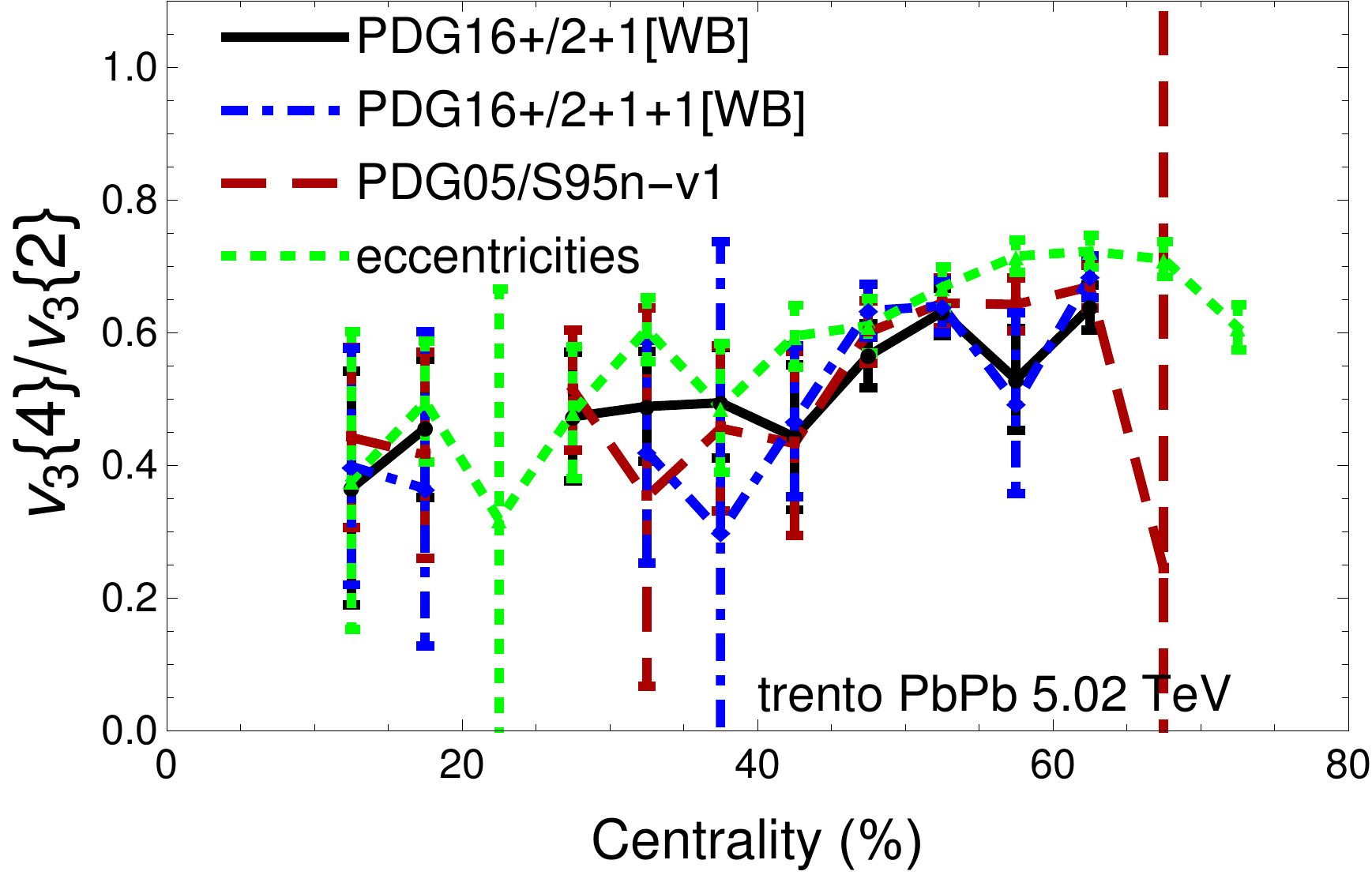} 
\end{tabular}
\caption{(Color online) $v_2\{4\}/v_2\{2\}$ (top) and $v_3\{4\}/v_3\{2\}$ (bottom) results for AuAu $\sqrt{s_{NN}}=200$ GeV collisions (left) and PbPb $\sqrt{s_{NN}}=5.02$ TeV collisions (right) for all charged particles computed using the S95n-v1 EoS from 2009 \cite{Huovinen:2009yb}, the 2+1 WB EoS from \cite{Borsanyi:2013bia}, and the 2+1+1 WB EoS from 2016 \cite{Borsanyi:2016ksw}. The green dashed line is the calculation using $\varepsilon_n\{4\}/\varepsilon_n\{2\}$.}
\label{fig:v2fluc}
\end{figure*}

In order to obtain the width of the distribution of flow fluctuations, normally the ratio of $v_n\{4\}/v_n\{2\}$ is used. In Fig.\ \ref{fig:v2fluc} $v_2\{4\}/v_2\{2\}$ (top) and $v_3\{4\}/v_3\{2\}$ (bottom) are shown for both RHIC and LHC run 2.  An advantage of this ratio between cumulants is that there are very small medium effects in central to mid-central collisions so information about the initial state eccentricities can be extracted directly.  Indeed, in Fig.\ \ref{fig:v2fluc} we see no effects from the choice of the equation of state, which confirms previous results \cite{Noronha-Hostler:2015dbi,Niemi:2015qia,Betz:2016ayq,Giacalone:2017uqx,Dusling:2017aot,Dusling:2017aot} that a measurement of $v_n\{4\}/v_n\{2\}$ is useful for investigating the properties of the initial state. 

In Fig. \ref{fig:v2fluc} there are gaps in the $v_3\{4\}/v_3\{2\}$ fluctuations.  This is because this is a very statistics-driven observable, especially in the more central collisions, so we would need many more events to obtain this calculation in certain bins. Finally, there does seem to be some slight energy dependence when comparing $v_2\{4\}/v_2\{2\}$ to $\varepsilon_2\{4\}/\varepsilon_2\{2\}$, which implies that the mapping from the initial state to the final flow harmonics likely also contains non-linear effects even in central collisions at RHIC.  It would be interesting to perform these comparisons in the RHIC Beam Energy Scan as well because, if this trend holds, the difference between $v_2\{4\}/v_2\{2\}$ and $\varepsilon_2\{4\}/\varepsilon_2\{2\}$ would become even more pronounced.

\begin{figure*}
\centering
\begin{tabular}{cc}
\includegraphics[width=0.5\textwidth]{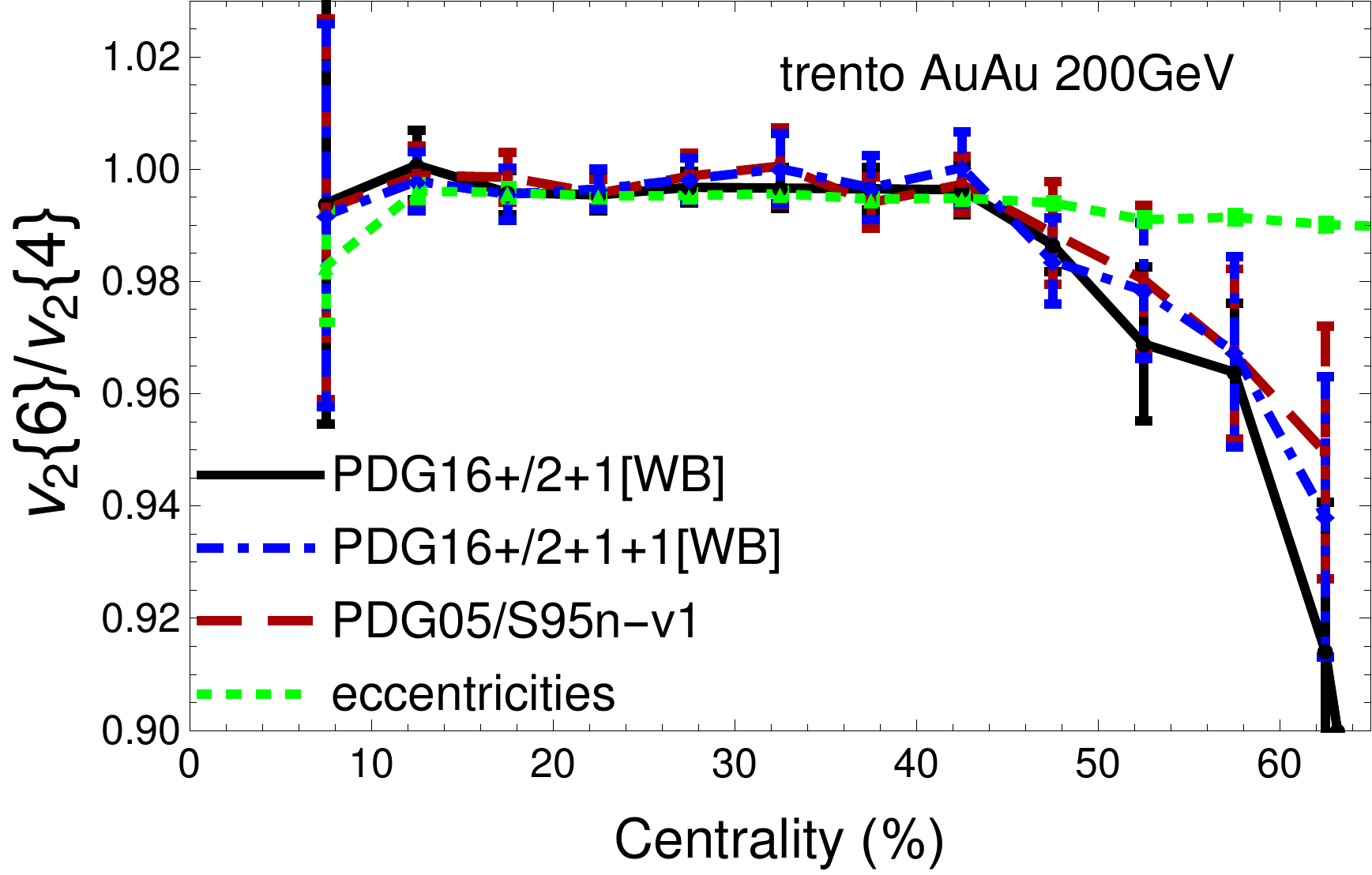} & \includegraphics[width=0.5\textwidth]{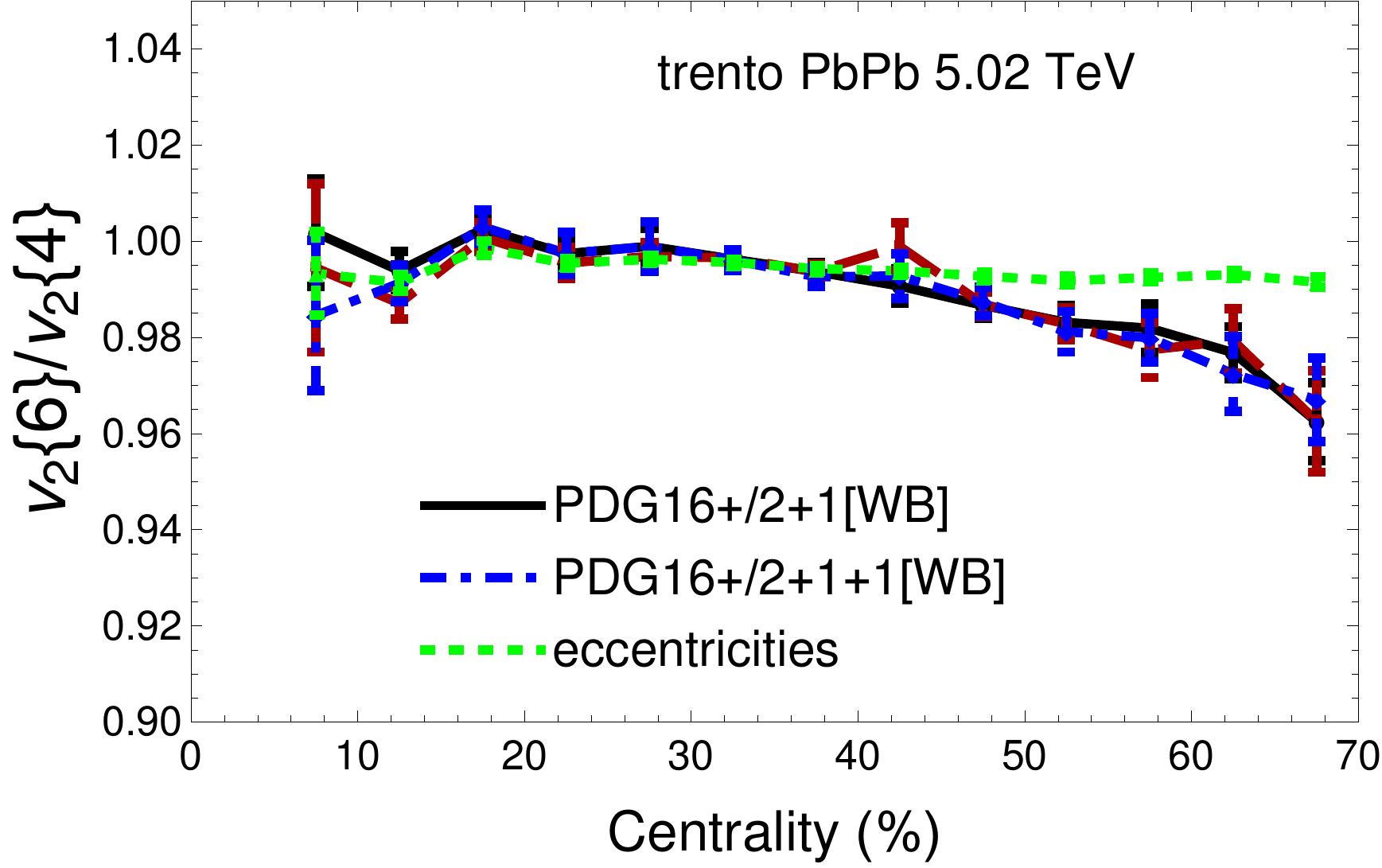} 
\end{tabular}
\caption{(Color online) $v_2\{6\}/v_2\{4\}$ results for AuAu $\sqrt{s_{NN}}=200$ GeV collisions (left) and PbPb $\sqrt{s_{NN}}=5.02$ TeV collisions (right) for all charged particles computed using the S95n-v1 EoS from 2009 \cite{Huovinen:2009yb}, the 2+1 WB EoS from \cite{Borsanyi:2013bia}, and the 2+1+1 WB EoS from 2016 \cite{Borsanyi:2016ksw}. The green dashed line is the calculation using $\varepsilon_2\{6\}/\varepsilon_2\{4\}$.}
\label{fig:v26fluc}
\end{figure*}

Higher order cumulants can also provide further insight into the skewness of the initial state \cite{Giacalone:2016eyu}. In Fig.\ \ref{fig:v26fluc} the ratio $v_2\{6\}/v_2\{4\}$ is shown, which can consistently be described by $\varepsilon_2\{6\}/\varepsilon_2\{4\}$ for $0-45\%$ centrality regardless of medium effects and energy. 

\begin{figure*}
\centering
\begin{tabular}{cc}
\includegraphics[width=0.5\textwidth]{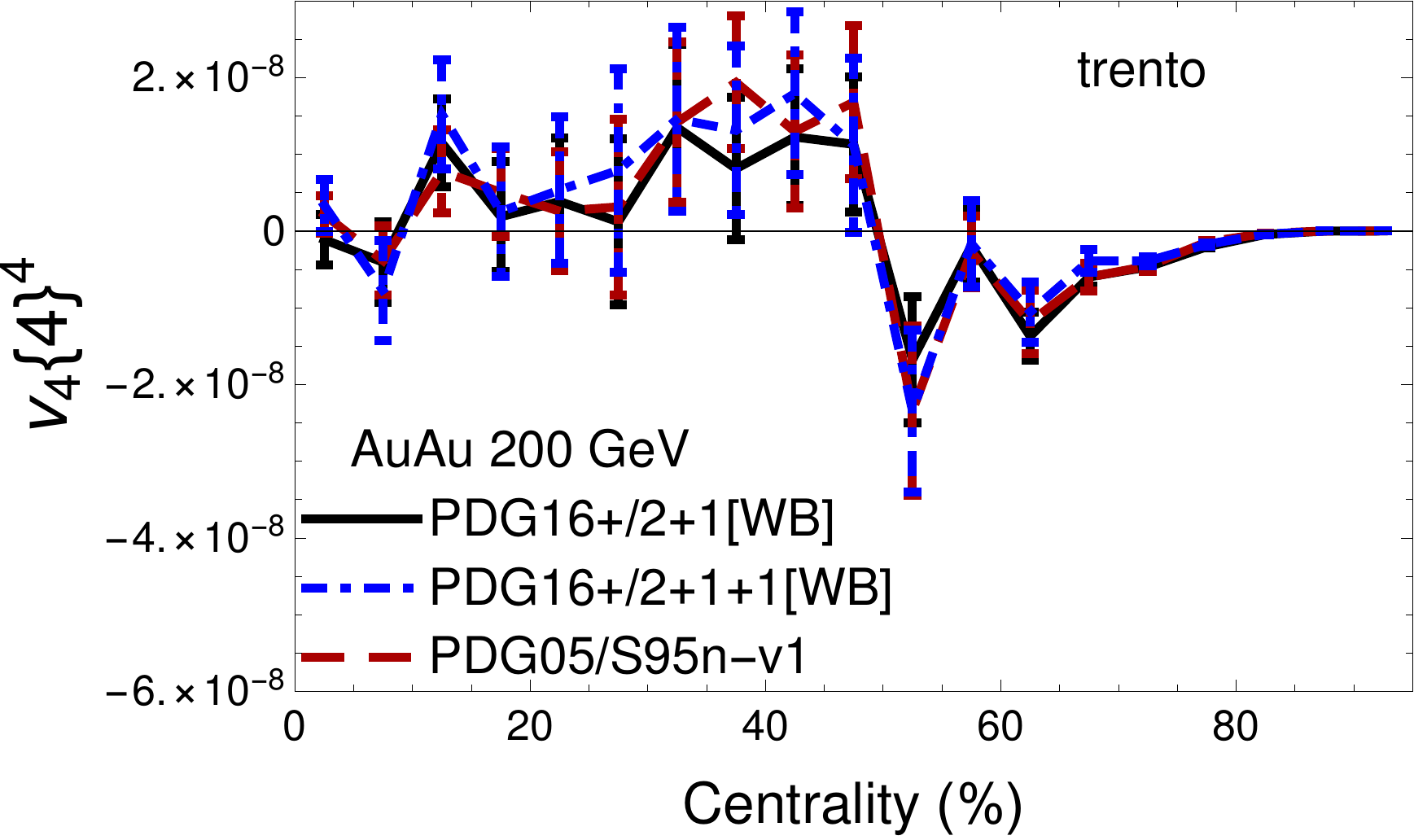} & \includegraphics[width=0.5\textwidth]{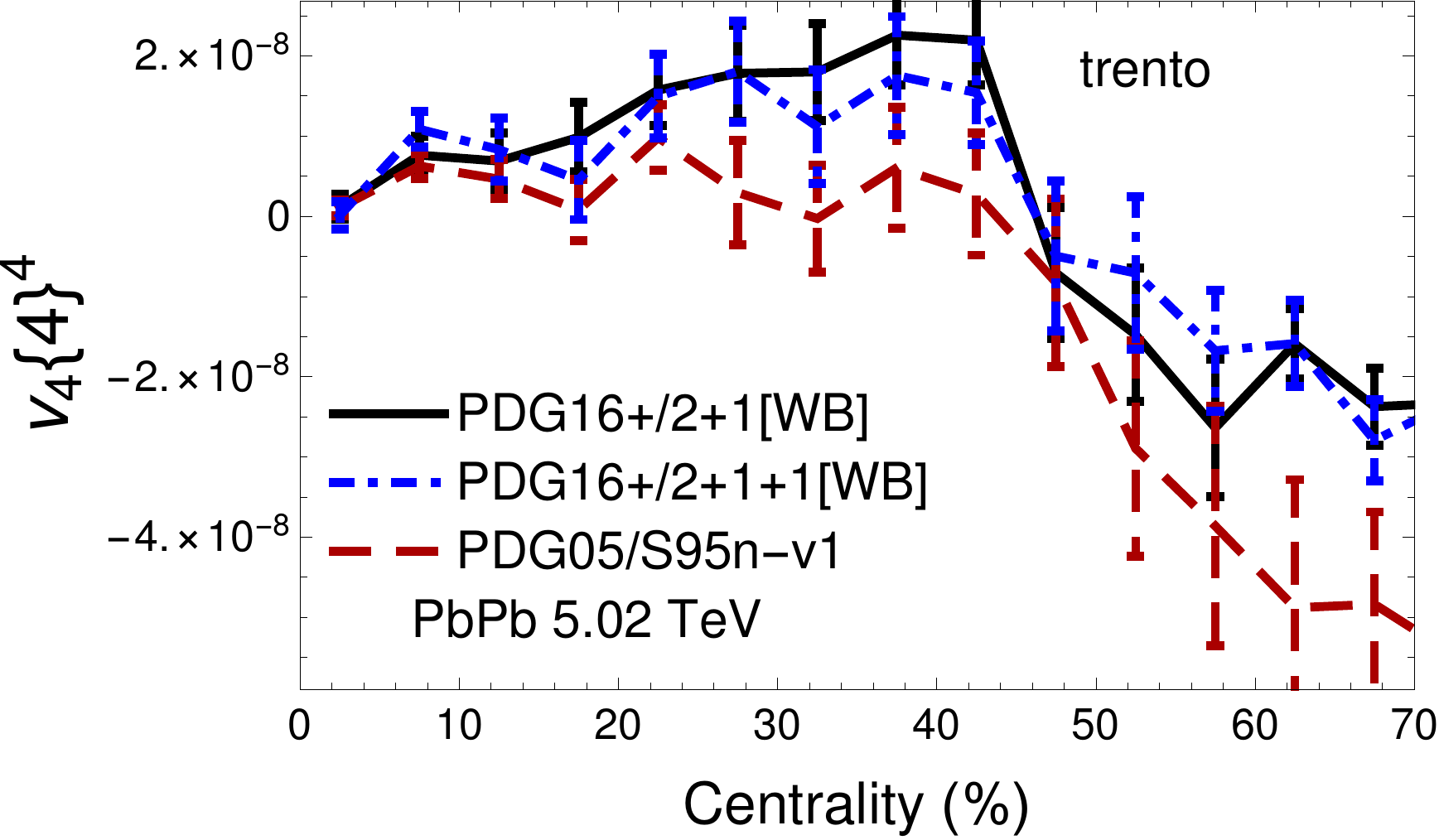}
\end{tabular}
\caption{(Color online) $(v_4\{4\})^4$ results for AuAu $\sqrt{s_{NN}}=200$ GeV collisions (left) and PbPb $\sqrt{s_{NN}}=5.02$ TeV collisions (right) for all charged particles computed using the S95n-v1 EoS from 2009 \cite{Huovinen:2009yb}, the 2+1 WB EoS from \cite{Borsanyi:2013bia}, and the 2+1+1 WB EoS from 2016 \cite{Borsanyi:2016ksw}. }
\label{fig:v44fluc}
\end{figure*}

Finally, in Fig.\ \ref{fig:v44fluc} $(v_4\{4\})^4$ is shown for both RHIC and LHC run 2.  Here we do not plot $v_4\{4\}$ directly because it changes sign \cite{Giacalone:2016mdr}, which would lead to imaginary numbers in central collisions.  As in \cite{Giacalone:2016mdr}, we see a sign change around $\sim 40\%$ centrality for both energies.  It does appear that this quantity depends on the choice of the equation of state at LHC run 2 though more statistics are needed to be certain. Interestingly enough, experiments also see a change in sign but this occurs closer to $20\%$ centrality \cite{Aamodt:2010pa} so a puzzle still remains. We note that this quantity has both linear and nonlinear contributions but the linear part is more dampened by viscosity so this could be used to constrain the temperature dependence of $\eta/s$ \cite{Giacalone:2016mdr}, although the high statistics needed for this observable requires significant computational resources that go beyond the scope of the present work. 

\begin{figure*}[ht]
\centering
\includegraphics[width=1\textwidth]{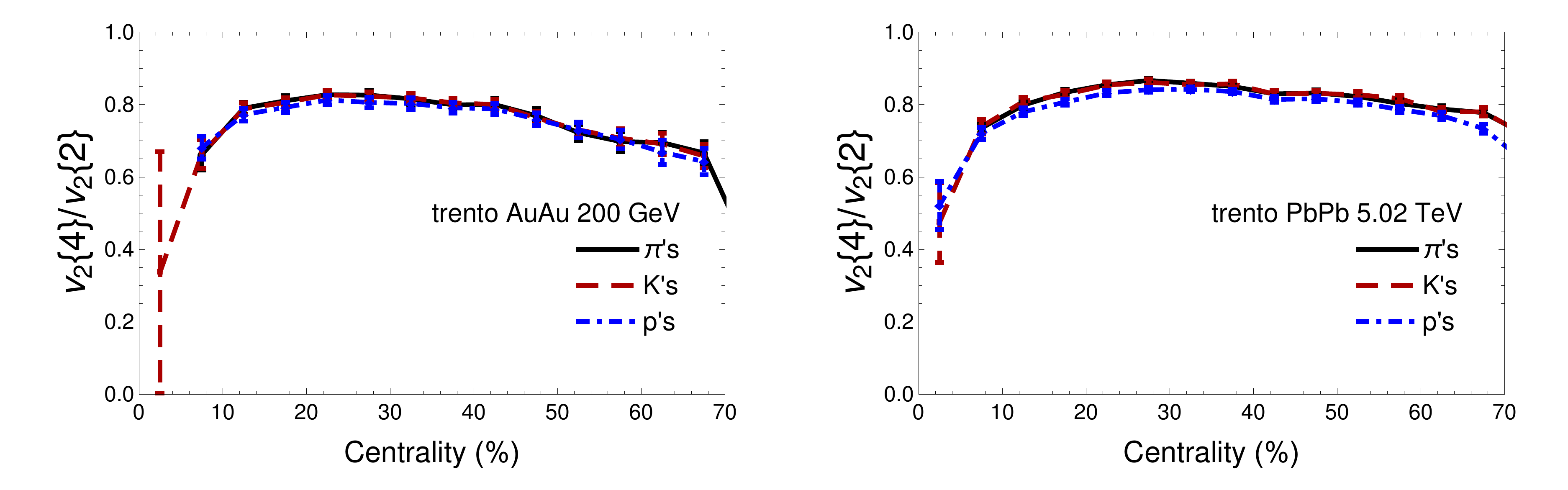}
\caption{(Color online)  $v_2\{4\}/v_2\{2\}$ for pions, kaons, and protons in AuAu $\sqrt{s_{NN}}=200$ GeV collisions (left) and PbPb $\sqrt{s_{NN}}=5.02$ TeV collisions (right) computed using the 2+1 WB EoS  \cite{Borsanyi:2013bia}.  }
\label{fig:cumid}
\end{figure*}
Also of interest is to look at flow fluctuations of identified particles.  Here we assume that there are high enough statistics to correlate 2 or 4 particles of interest since we restrict our study to $\pi$'s, p's, and K's.  Thus, the cumulants are still described using Eqs.\ (\ref{eqn:cumulants}) but with identified particles instead of all charged particles.  In Fig.\ \ref{fig:cumid}  the results are shown for $v_2\{4\}/v_2\{2\}$ for the  2+1 WB EoS  \cite{Borsanyi:2013bia}\footnote{Since there is no equation of state dependence for this observable we see no reason to present the results from other choices.}. While non-linear effects can play a role in the magnitude of $v_2\{2\}$ for identified particles  (see Fig.\ \ref{fig:q2pid}), all effects cancel out for the ratio $v_2\{4\}/v_2\{2\}$.  This strengthens the arguments that $v_2\{4\}/v_2\{2\}$ arises from initial state effects.

\subsection{Symmetric cumulants}\label{sec:sc}

The event-by-event correlations among fluctuations of flow harmonics of different order encode both information about the medium and initial state \cite{ALICE:2016kpq,Zhou:2016eiz,Zhu:2016puf,Gardim:2016nrr,Giacalone:2016afq,PhysRevC.95.044902,Ke:2016jrd,Eskola:2017imo}. To study this, symmetric cumulants are used (here we only considered normalized symmetric cumulants):
\begin{equation}\label{eqn:NSC}
NSC(m,n)=\frac{\langle v_m^2 v_n^2\rangle-\langle v_m^2\rangle\langle v_n^2\rangle}{\langle v_m^2\rangle\langle v_n^2\rangle}
\end{equation}
where $m\neq n$.  For all observables multiplicity weighing and centrality rebinning are included to avoid artificial centrality dependences \cite{Gardim:2016nrr}. Note that Eq.\ (\ref{eqn:NSC}) can be calculated with the eccentricities alone, which will be denoted as $\varepsilon SC(m,n)$ throughout this paper.

\begin{figure*}[ht]
\centering
\includegraphics[width=1\textwidth]{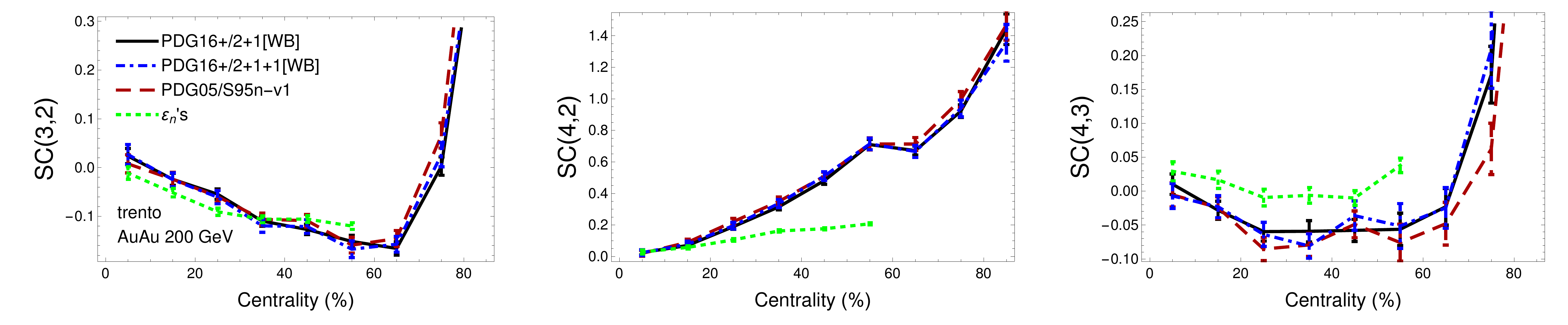} \\
\includegraphics[width=1\textwidth]{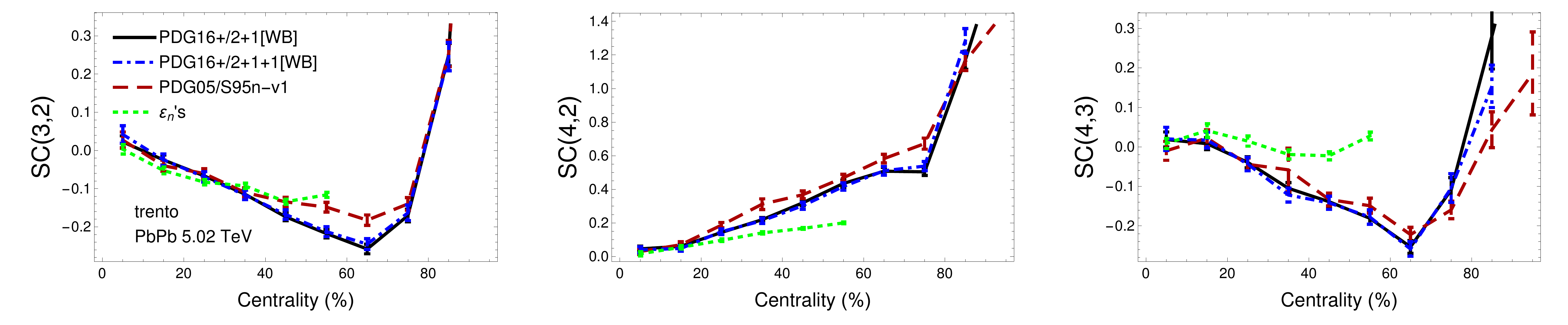} 
\caption{(Color online) Symmetric cumulants results for all charged particles in AuAu $\sqrt{s_{NN}}=200$ GeV collisions (top) and PbPb $\sqrt{s_{NN}}=5.02$ TeV collisions (bottom) computed using the S95n-v1 EoS from 2009 \cite{Huovinen:2009yb}, the 2+1 WB EoS from \cite{Borsanyi:2013bia}, and the 2+1+1 WB EoS from 2016 \cite{Borsanyi:2016ksw}. The green dashed line is the calculation using $\varepsilon SC(m,n)$.}
\label{fig:SC}
\end{figure*}

Symmetric cumulants have been shown to be a promising observable for testing the initial state and $NSC(3,2)$, especially appears to be independent of viscous effects \cite{Gardim:2016nrr}. Meanwhile, other combinations, such as $NSC(4,2)$, appear to include non-linear effects driven by viscous effects.  One should also note that the symmetric cumulants can be correlated to the event plane correlations \cite{Giacalone:2016afq}.  In Fig.\ \ref{fig:SC} the symmetric cumulants are shown for $NSC(3,2)$, $NSC(4,2)$, and $NSC(4,3)$.  

There does appear to be some slight dependence of the symmetric cumulants on the equation of state for LHC run 2 but only for the comparison between the old equation of state vs. the two newest ones from the WB collaboration. However, even those differences only show up for peripheral collisions where the error bars are larger, which implies that symmetric cumulants are not the best observable for distinguishing between different assumptions regarding the QCD equation of state used in hydrodynamic simulations.

\begin{figure*}[ht]
\centering
\includegraphics[width=1\textwidth]{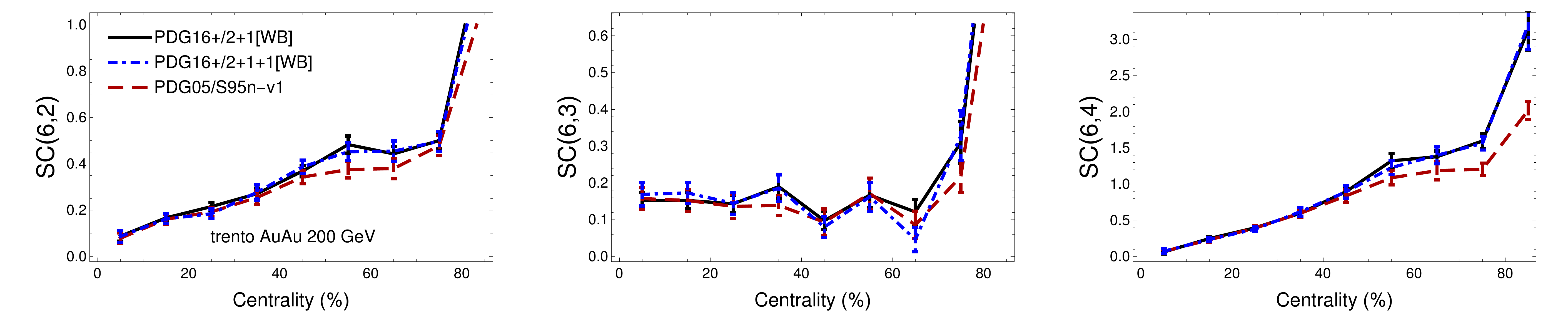} \\
\includegraphics[width=1\textwidth]{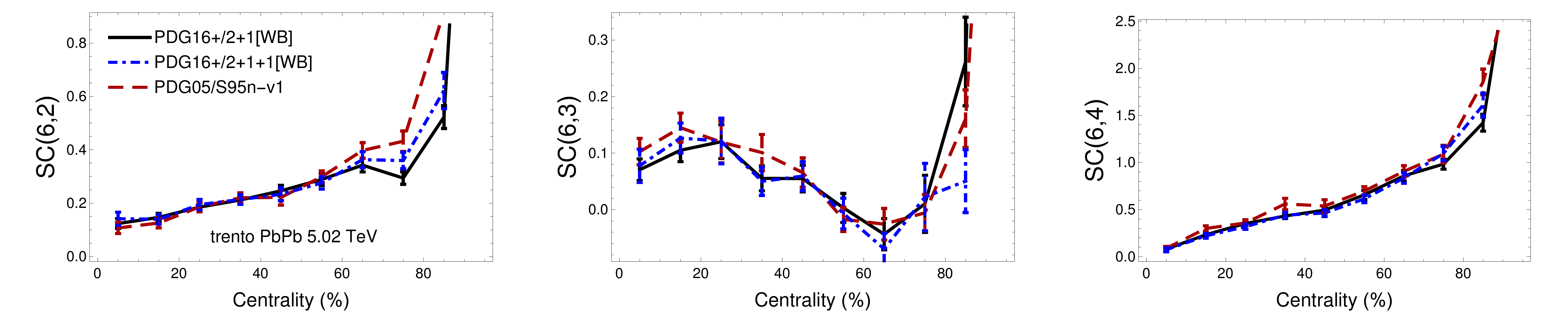} 
\caption{(Color online) Symmetric cumulants results with $v_6$ combinations for all charged particles in AuAu $\sqrt{s_{NN}}=200$ GeV collisions (top) and PbPb $\sqrt{s_{NN}}=5.02$ TeV collisions (bottom) computed using the S95n-v1 EoS from 2009 \cite{Huovinen:2009yb}, the 2+1 WB EoS from \cite{Borsanyi:2013bia}, and the 2+1+1 WB EoS from 2016 \cite{Borsanyi:2016ksw}.}
\label{fig:SC6}
\end{figure*}

In Fig.\ \ref{fig:SC6} symmetric cumulants with $v_6$ are included for $NSC(6,2)$, $NSC(6,3)$, and $NSC(6,4)$. Previously, it was found that there is a certain degree of numerical error involved with $v_6$ calculations from v-USPhydro \cite{Noronha-Hostler:2013gga}, however, the error has not yet been studied with viscous hydrodynamics and integrated flow harmonics. Still, we are motivated to study symmetric cumulants of $v_6$ with the hope that there might be some hints of influence of the equation of state.  We find that RHIC sees large correlations between $v_6$ and other flow harmonics compared to LHC. Additionally, $v_6$ appears to be the most strongly correlated with $v_4$ and to a lesser extent with $v_2$. One can see a very slight dependence on the different equations of state for peripheral collisions but none of these are large enough to make these quantities strong candidates for distinguishing between different assumptions that go into the equation of state.

\begin{figure*}[ht]
\centering
\includegraphics[width=1\textwidth]{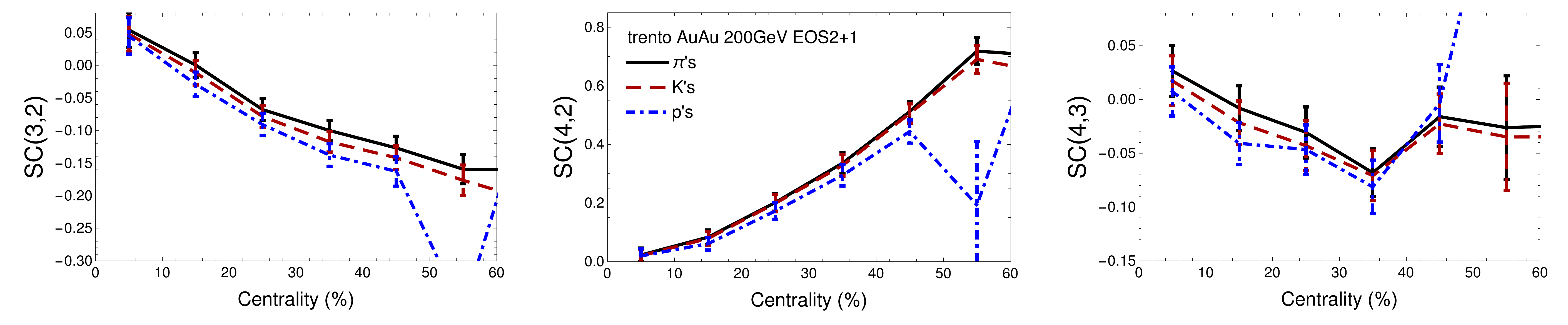} \\
\includegraphics[width=1\textwidth]{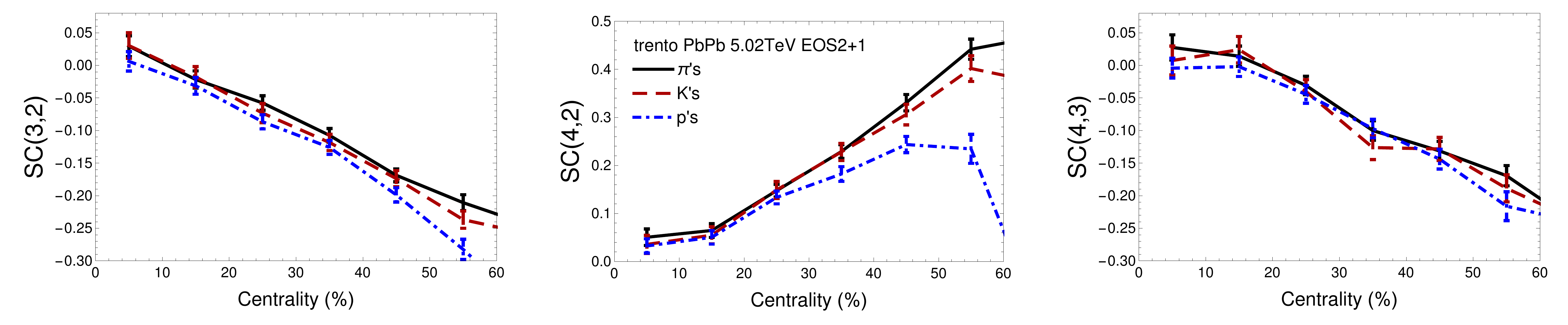} 
\caption{(Color online) Symmetric cumulants results for pions, kaons, and protons in AuAu $\sqrt{s_{NN}}=200$ GeV collisions (top) and PbPb $\sqrt{s_{NN}}=5.02$ TeV collisions (bottom) computed using the 2+1 WB EoS  \cite{Borsanyi:2013bia}.}
\label{fig:SCid}
\end{figure*}
In Fig.\ \ref{fig:SCid} the symmetric cumulants $NSC(3,2)$, $NSC(4,2)$, and $NSC(4,3)$ for $\pi$'s, K's, and p's, computed using the 2+1 WB EoS  \cite{Borsanyi:2013bia}, are presented for both RHIC (top) and LHC run 2 (bottom).  We find that these quantities show some dependence on the mass of the identified particle.  For instance,  $NSC(3,2)$ is {\it more} anti-correlated for heavier particles whereas  $NSC(4,2)$ is {\it less} correlated for protons.  Calculating $NSC(m,n)$ in peripheral collisions requires higher statistics for protons so we leave a deeper analysis for a future paper where we can analyze a larger set of events. It would be interesting to have experimental results for the symmetric cumulants by particle ID in order to verify this mass scaling effect.

\subsection{$\sqrt{s_{NN}}$ dependence}

The difference between flow harmonics measured at RHIC and LHC run 1 have been studied in \cite{Abreu:2007kv} while in \cite{Noronha-Hostler:2015uye,Niemi:2015voa,McDonald:2016vlt} the corresponding differences between LHC run 1 and LHC run 2 were investigated. However, the largest difference between collision energies where the assumption that $\mu_B\sim 0$ holds is between AuAu collisions at $\sqrt{s_{NN}}=200$ GeV and PbPb collisions at $\sqrt{s_{NN}}=5.02$ TeV.  How high-statistics observables such as multiparticle cumulants and symmetric cumulants scale with collision energy has not been studied in depth in hydrodynamic models and, as a matter of fact, in the case of symmetric cumulants these quantities have not been measured at all the energies yet.   

\begin{figure*}[ht]
\centering
\includegraphics[width=\textwidth]{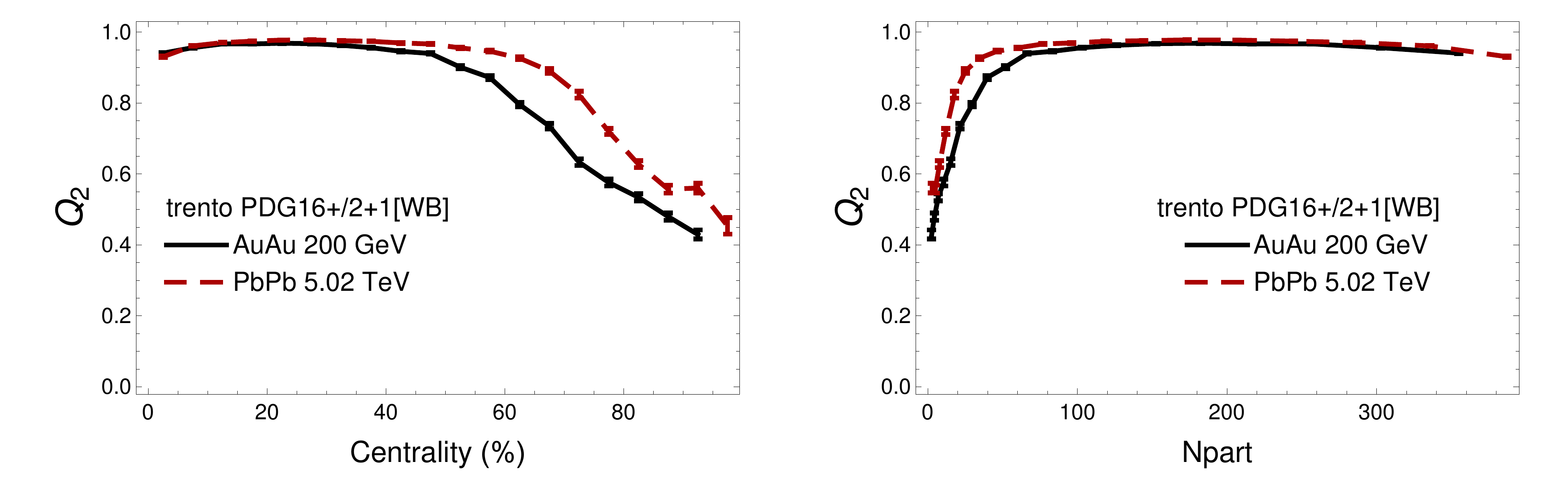} 
\caption{(Color online) Pearson coefficient for the linear mapping between $\varepsilon_2$ and $ v_2$ for all charged particles in AuAu $\sqrt{s_{NN}}=200$ GeV collisions compared to PbPb $\sqrt{s_{NN}}=5.02$ TeV collisions computed using the 2+1 WB EoS from \cite{Borsanyi:2013bia}. Here we scale both by the centrality (left) and the number of participants (right). }
\label{fig:q2snn}
\end{figure*}

Before we make comparisons between experimental observables across different beam energies, we first examine the basic linear mapping between $\varepsilon_2$ and $ v_2$ for all charged particles when going from AuAu $\sqrt{s_{NN}}=200$ GeV to PbPb $\sqrt{s_{NN}}=5.02$ TeV collisions.  In all the following results we only take the  2+1 WB EoS from \cite{Borsanyi:2013bia} because we did not see a strong equation of state dependence on these observables.  In Fig.\ \ref{fig:q2snn} the Pearson coefficients are shown across beam energies either scaled by centrality or the number of participants Npart.  In both cases it is clear that at RHIC non-linear effects play a larger role, especially in peripheral collisions.  This is likely one of the reasons why constraints for the temperature dependence of $\eta/s$ have been more successful at RHIC energies \cite{Adamczyk:2017byf}. However, this does imply that it may be much more difficult to probe the high temperature regime of $\eta/s$ \cite{Niemi:2012ry}. We note that the particle identification effects found here appear to be of relatively equal magnitudes for both beam energies so a study at either RHIC or LHC should provide similar information. 

\begin{figure*}[ht]
\centering
\includegraphics[width=\textwidth]{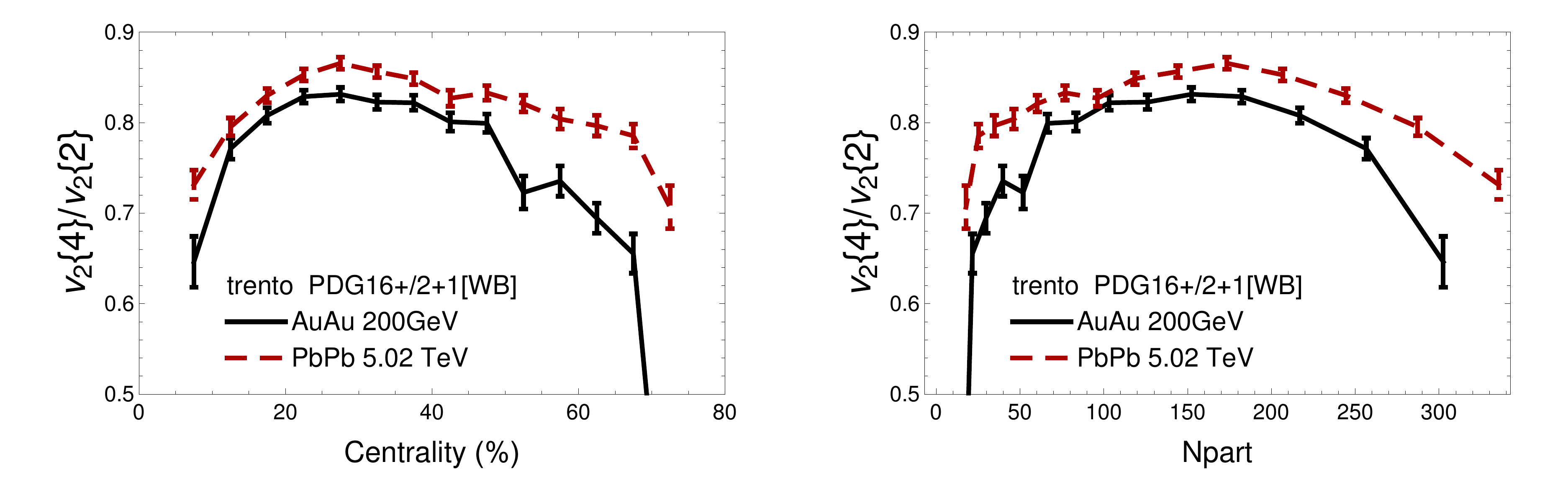} 
\caption{(Color online) $v_2\{4\}/v_2\{2\}$ results for all charged particles in AuAu $\sqrt{s_{NN}}=200$ GeV collisions compared to PbPb $\sqrt{s_{NN}}=5.02$ TeV collisions computed using the 2+1 WB EoS from \cite{Borsanyi:2013bia}. Here we scale both by the centrality (left) and the number of participants (right).}
\label{fig:v24v22snn}
\end{figure*}

In Fig.\ \ref{fig:v24v22snn} we see  that  $v_2\{4\}/v_2\{2\}$ increases with the beam energy.  This implies that LHC run 2 energies experience less $v_2$ fluctuations (so one would expect a narrower $v_2$ distribution).  When scaling by both Npart and centrality this splitting between the two energies remains. Since it was already shown in Figs.\ \ref{fig:v2fluc} that $v_2\{4\}/v_2\{2\}$  does not exhibit any medium effects, this behavior arises directly from the initial conditions themselves.

\begin{figure}[ht]
\centering
\includegraphics[width=0.5\textwidth]{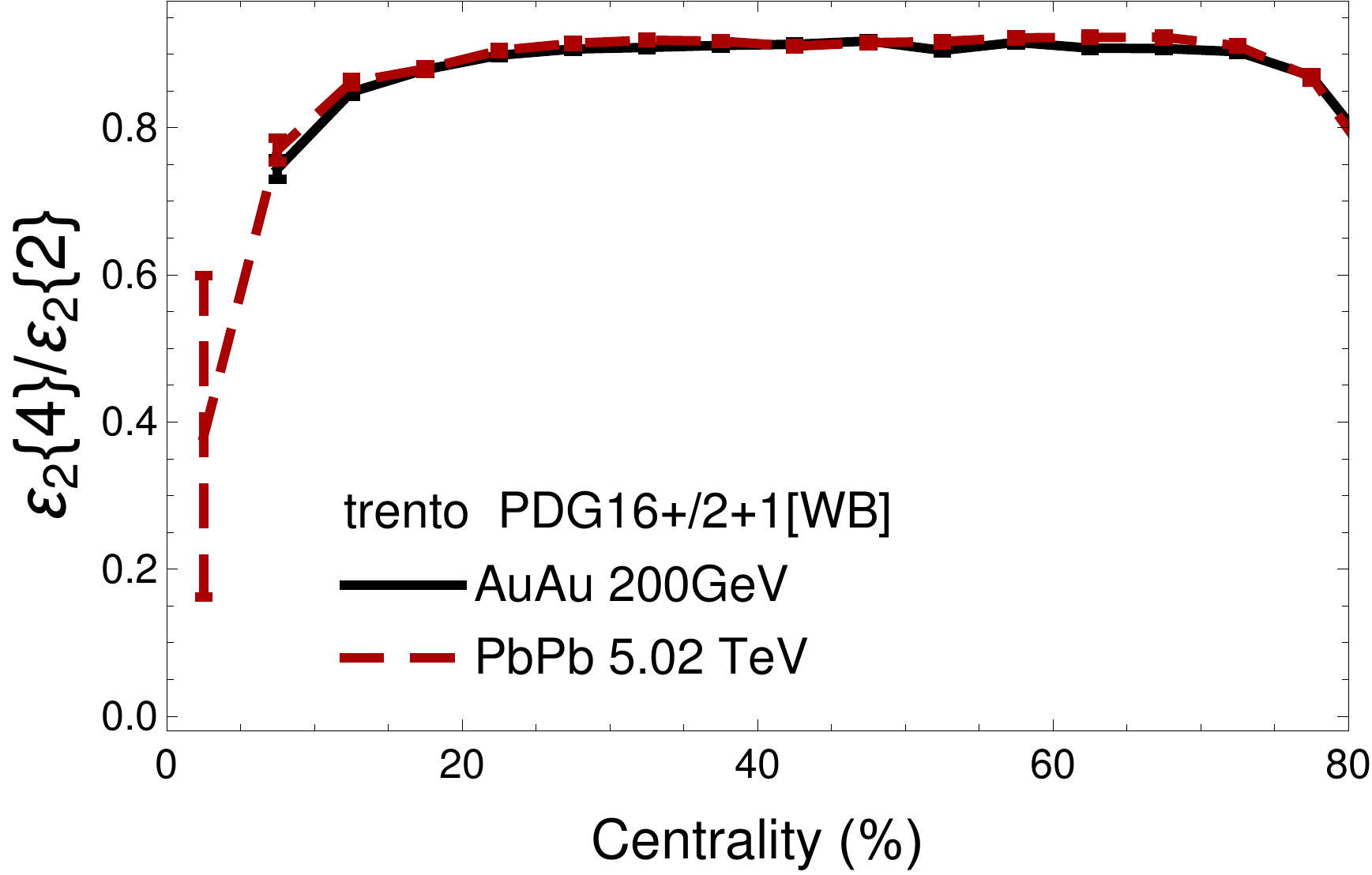} 
\caption{(Color online) $\varepsilon_2\{4\}/\varepsilon_2\{2\}$ results for AuAu $\sqrt{s_{NN}}=200$ GeV collisions compared to PbPb $\sqrt{s_{NN}}=5.02$ TeV collisions.}
\label{fig:e24v22snn}
\end{figure}

In order to test how much the initial conditions play a role in $v_2\{4\}/v_2\{2\}$ at different $\sqrt{s_{NN}}$, we plot $\varepsilon_2\{4\}/\varepsilon_2\{2\}$ results for AuAu $\sqrt{s_{NN}}=200$ GeV collisions compared to PbPb $\sqrt{s_{NN}}=5.02$ TeV collisions in Fig.\ \ref{fig:e24v22snn}. Surprisingly, there is no dependence of $\varepsilon_2\{4\}/\varepsilon_2\{2\}$ on the beam energy.  This implies that the highest LHC energy is our best bet for constraining initial conditions via $v_n$ fluctuations (if one wants to directly compare $v_2\{4\}/v_2\{2\}$ with $\varepsilon_2\{4\}/\varepsilon_2\{2\}$).  Otherwise, at RHIC energies one must run the full hydrodynamic simulation to determine $v_2\{4\}/v_2\{2\}$ even though medium effects from $\eta/s$ and EoS are not apparent.

\begin{figure*}[ht]
\centering
\includegraphics[width=1\textwidth]{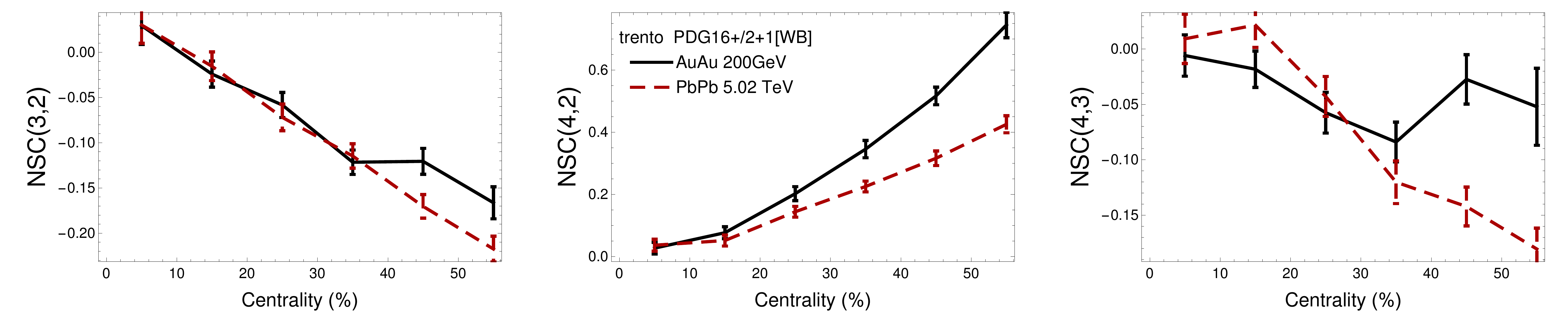} \\
\includegraphics[width=1\textwidth]{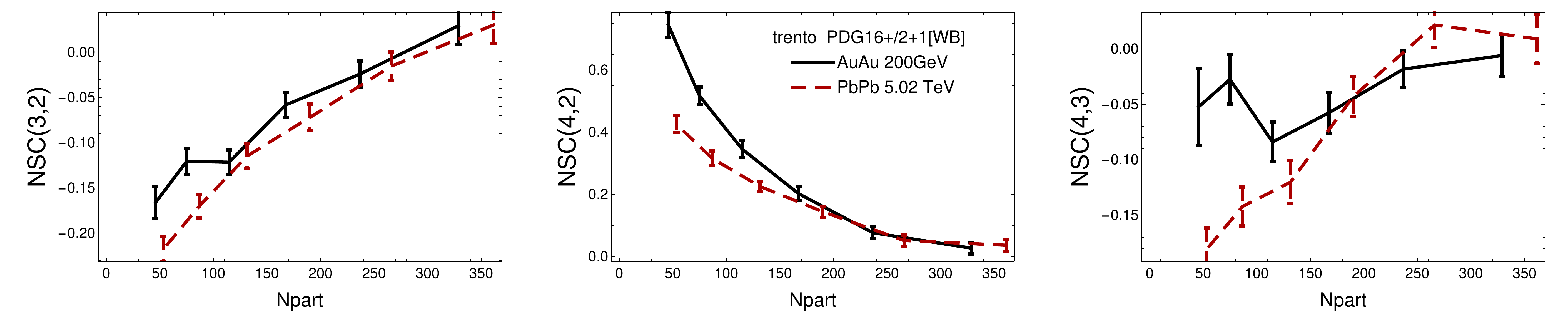}
\caption{(Color online) Symmetric cumulant results for all charged particles in AuAu $\sqrt{s_{NN}}=200$ GeV collisions compared to PbPb $\sqrt{s_{NN}}=5.02$ TeV  collisions computed using the 2+1 WB EoS from \cite{Borsanyi:2013bia}. Here we scale both by the centrality (top) and the number of participants (bottom).}
\label{fig:scsnn}
\end{figure*}

In Fig.\ \ref{fig:scsnn} the symmetric cumulants $NSC(3,2)$, $NSC(4,2)$, and $NSC(4,3)$ computed at two different energies are shown as functions of centrality (top) and Npart (bottom). We find that $NSC(3,2)$ is almost identical across energies as a function of centrality (below $40\%$) whereas a strong anti-correlation for larger energies is obtained when this quantity is plotted as a function of Npart. The reverse can be said about $NSC(4,2)$.  For both types of scalings, one expects that lower energies see a larger correlation between $v_2$ and $v_4$. However, that difference is largest when scaled by the centrality whereas for Npart scaling it disappears in central to mid-central collisions and only appears in peripheral collisions.  Because we already showed in Fig.\ \ref{fig:q2snn} that non-linear effects play a larger role in $v_2$ at RHIC, this increase in   $NSC(4,2)$ can likely be attributed to that enhancement in non-linear effects. The role of non-linear effects, especially at RHIC, can also be seen in Fig.\ \ref{fig:SC} by comparing $NSC(4,2)$ calculated using $v_n$'s to the corresponding estimate for this quantity obtained using only the eccentricities $\varepsilon_n$'s.

The underlying cause of this beam energy scaling is not very clear.  In the case of the 2+1 WB EoS, we switch on hydrodynamics at the same time at RHIC and LHC run 2, and we also freeze-out at the same temperature.  However, because LHC reaches higher temperatures, the system lives longer at LHC run 2 than at RHIC so the dependence with the beam energy may be due to different lifetimes of the QGP in these systems.  Additionally, the (average) initial conditions are certainly smaller in radius for AuAu vs. PbPb so the system size could also be playing a role. We plan on exploring these differences in more detail in a future paper. 

\section{Conclusions}\label{sec:con}

In this paper we constructed two new equations of state using the state-of-the-art lattice QCD calculations for 2+1 and 2+1+1 quark flavors combined with all the PDG resonances that have been shown to be relevant for partial pressure calculations in comparison to lattice QCD calculations. The decay channels of these new resonances were either taken directly from experimental data or extrapolated from existing information from neighboring particles with the same quantum statistics. We then studied the effects of the different equations of state and the new resonances on flow observables, focusing only on integrated quantities. We have also made predictions for observables that could be measured at LHC run 2 and also at RHIC.  

Regarding the inclusion of new resonances, we found that their influence primarily dominates in the spectra where they enhance the number of high $p_T$ particles, which leads to a larger $\langle p_T\rangle$.  Additionally, there is a significant improvement in the description of protons due to the extra resonances.  This appears to be relatively universal with similar effects at both RHIC and at LHC run 2.

In this paper, the largest influence of the equation of state is in the extraction of $\eta/s$.  At RHIC energies, lower temperatures are reached where the differences in the equations of states are smaller so the same $\eta/s$ can be used for all three equations of state, which is in line with the Bayesian analysis in Ref.\ \cite{Bernhard:2016tnd} and a study using ideal hydrodynamics \cite{Dudek:2014qoa}.  At LHC run 2 larger temperatures around $T\sim 600$ MeV can be reached (although these values depend on the individual equation of state) and each different equation of state corresponds to a different $\eta/s$. We find that the commonly used S95n-v1 requires roughly half the viscosity as the new  PDG16+/2+1(+1)[WB] equations of state.  Even between the 2+1 and 2+1+1 flavor equations of state there is roughly a $15\%$ difference in $\eta/s$ at LHC run 2.  Therefore, we cannot yet claim to have solid evidence that the contribution from charm quarks needs to be included in the EoS used in hydrodynamic simulations. These new equations of state combined with the updated particle resonance decays will be made publicly available for the scientific community soon on \url{https://github.com/jnoronhahostler/Equation-of-State}.

While the assumptions regarding the equation of state affect $\eta/s$, those make little to almost no difference on most of the observables studied in this paper.  The observables that have any visible difference (albeit small) are the $\langle p_T\rangle$ of identified particles at LHC run 2 and peripheral flow harmonics at LHC run 2 (only differences between S95n-v1 vs. PDG16+/2+1(+1)[WB] seen). Two other observables provide possible hints of differences between equations of state:  $v_2\{2\}/v_3\{2\}$ in ultra central collisions and $v_4\{4\}^4$ both at LHC run 2. However, conclusions about these high-statistics observables in this regard would require many more events than the 30,000 used in this paper. In the case of $v_2$ fluctuations, i.e, $v_2\{4\}/v_2\{2\}$ and $v_2\{6\}/v_2\{4\}$, we found absolutely no dependence on the equation of state, which continues to demonstrate that fluctuation observables are likely to give our best constraints on initial conditions \cite{Giacalone:2017uqx}.

We also point out that we find some discrepancies with respect to the data for certain observables, which may provide room for determining $\eta/s$ as a function of temperature.  For instance, our $v_3\{2\}$ at RHIC is slightly too large and, as was shown in \cite{Niemi:2015qia}, it is strongly dependent on $\eta/s(T)$ there.  For peripheral collisions at both RHIC and LHC run 2 we find that our flow harmonics converge to zero. However, this may also be due to non-flow effects that will be removed in the future when sub-events are implemented \cite{Jia:2017hbm,Aaboud:2017blb,DiFrancesco:2016srj}. At RHIC energies we find that non-linear effects are more important, which is likely why Ref. \cite{Niemi:2015qia} found a strong sensitivity to $\eta/s$ as a function of the temperature at RHIC energies.  Here we showed that the Pearson coefficient between $\varepsilon_2$ and $v_2$ is smaller at RHIC energies and that there is a larger deviation between $SC(4,2)$ and $\varepsilon SC(4,2)$ computed using only the eccentricities, both of which are strong indications of non-linear effects in $v_2$.  

Because of the inclusion of new hadronic resonances, we were interested in the effects on flow observables of identified particles.  We found that non-linearities are more relevant in heavier particles and we wanted to explore the consequences of this effect on new observables (beyond the standard two particle correlations).  Again looking at $v_2$ fluctuations we found no effect by particle identification, which indicates once again that these fluctuations originate from the initial state. However, symmetric cumulants do appear to have a dependence on the mass of the identified particles such that protons see a larger anti-correlation between $v_2$ and $v_3$ than pions whereas $v_2$ and $v_4$ are less correlated for protons compared to pions. We hope that experimentalists measure symmetric cumulants of identified particles to see whether this mass ordering is observed experimentally.

Finally, we have studied how observables scale when going from RHIC $200$ GeV to LHC $5.02$ TeV.  One particularly interesting finding is that the ratio $v_2\{2\}/v_3\{2\}$ gets closer to unity as the beam energy is increased.  This may be an indication that hydrodynamic simulations need to be run for longer periods of time i.e. a smaller $\tau_0$ and/or lower $T_{FO}$ in order to see the convergence of $v_2\{2\}/v_3\{2\}\rightarrow 1$ in ultracentral collisions.  For $0-2\%$ centrality we see the closest match to data using the 2+1 WB EoS but our result still remains about $15\%$ above the data.  In order to understand this puzzle in ultracentral collisions it would be very useful for experimentalists to measure $v_2\{2\}/v_3\{2\}$ across different beam energies to see if they see the same general trend obtained in our calculations.  

We also studied how fluctuations scale across energies and found that $v_2\{4\}/v_2\{2\}$ decreases with increasing $\sqrt{s_{NN}}$, which implies that RHIC has a wider $v_2$ distribution than LHC run 2. It is interesting to note that at RHIC there is a slightly larger deviation between $\varepsilon_2\{4\}/\varepsilon_2\{2\}$ and $v_2\{4\}/v_\{2\}$. We expect that for the Beam Energy Scan this deviation would continue to increase, which would be another interesting measurement that could be done at RHIC.

\section*{Acknowledgements}
The authors would like to thank Anthony Timmins, Ron Belmont, Jamie Nagle, You Zhou, Szabolcs Borsanyi, Matthew Luzum, Jean-Yves Ollitrault, Giacalone Giuliano, Chun Shen, and Bjoern Schenke for discussions related to this work. 
J.N. thanks Conselho Nacional de Desenvolvimento Cient\'{\i}fico e
Tecnol\'{o}gico (CNPq) and Funda\c c\~ao de
Amparo \`{a} Pesquisa do Estado de S\~{a}o Paulo (FAPESP) under grant
2015/50266-2 for financial support and the Department of Physics and
Astronomy at Rutgers University for its hospitality.
J.N.H. acknowledges the Office of Advanced Research Computing (OARC) at Rutgers, The State University of New Jersey for providing access to the Amarel cluster and associated research computing resources that have contributed to the results reported here. J.N.H also acknowledges the use of the Maxwell Cluster and the advanced support from the Center of Advanced Computing and Data Systems at the University of Houston.   This material is based upon work supported by the National Science Foundation under grants no. PHY-1654219 and OAC-1531814 and by the U.S. Department of Energy, Office of Science, Office of Nuclear Physics, within the framework of the Beam Energy Scan Theory (BEST) Topical Collaboration. 

\section*{References}
\bibliography{all}

\end{document}